\documentclass[twocolumn]{aastex63}

\def\a{\alpha}
\def\b{\beta}
\def\g{\gamma}
\def\y{\gamma}
\def\d{\delta}
\def\D{\Delta}
\def\e{\epsilon}
\def\i{\iota}
\def\k{\kappa}
\def\l{\lambda}
\def\u{\mu}
\def\v{\nu}
\def\s{\sigma}
\def\t{\tau}

\def\x{\xi}

\def\n{\nabla}
\def\w{\omega}
\def\z{\zeta}

\def\p{\phi}

\def\L{\Lambda}

\def\O{\Omega}
\usepackage{lipsum}

\shorttitle{\textsc{A Link Between White Dwarf Pulsars and Polars}}
\shortauthors{Rodriguez et al.}
\graphicspath{{./}{figures/}}

\usepackage{float}
\usepackage{appendix}
\usepackage{xcolor}
\usepackage{amsmath}
\usepackage[T1]{fontenc}

\begin{document}


\title{A Link Between White Dwarf Pulsars and Polars: Multiwavelength Observations of the 9.36-Minute Period Variable Gaia22ayj}

\correspondingauthor{Antonio C. Rodriguez}
\email{acrodrig@caltech.edu}

\author[0000-0003-4189-9668]{Antonio C. Rodriguez}
\affiliation{Department of Astronomy, California Institute of Technology, 1200 East California Blvd, Pasadena, CA, 91125, USA}

\author[0000-0002-6871-1752]{Kareem El-Badry}
\affiliation{Department of Astronomy, California Institute of Technology, 1200 East California Blvd, Pasadena, CA, 91125, USA}

\author{Pasi Hakala}
\affiliation{Finnish Centre for Astronomy with ESO (FINCA), Quantum, University of Turku, FI-20014, Finland}

\author[0000-0002-4717-5102]{Pablo Rodríguez-Gil}
\affiliation{Instituto de Astrofísica de Canarias, E-38205 La Laguna, Tenerife, Spain} 
\affiliation{Departamento de Astrofísica, Universidad de La Laguna, E-38206 La Laguna, Tenerife, Spain}

\author[0000-0002-5082-5049]{Tong Bao}
\affiliation{INAF – Osservatorio Astronomico di Brera, Via E. Bianchi 46, 23807 Merate (LC), Italy}

\author[0000-0001-5778-2355]{Ilkham Galiullin}
\affiliation{Kazan Federal University, Kremlevskaya Str.18, 420008, Kazan, Russia}

\author[0009-0005-5452-0671]{Jacob A. Kurlander}
\affiliation{Department of Astronomy, University of Washington, 3910 15th Avenue NE, Seattle, WA 98195, USA}

\author[0000-0002-4119-9963]{Casey J. Law}
\affiliation{Department of Astronomy, California Institute of Technology, 1200 East California Blvd, Pasadena, CA, 91125, USA}
\affiliation{Owens Valley Radio Observatory, California Institute of Technology, Big Pine CA 93513, USA}

\author[0000-0003-4615-6556]{Ingrid Pelisoli}
\affiliation{Department of Physics, University of Warwick, Coventry CV4 7AL, UK}

\author[0000-0003-3903-8009]{Matthias R. Schreiber}
\affiliation{Departamento de F{\'i}sica, Universidad T{\'e}cnica Federico Santa Mar{\'i}a, Av. Espa{\~n}a 1680, Valpara{\'i}so, Chile}

\author[0000-0002-7226-836X]{Kevin Burdge}
\affiliation{Department of Physics, Massachusetts Institute of Technology, Cambridge, MA, USA}
\affiliation{Kavli Institute for Astrophysics and Space Research, Massachusetts Institute of Technology, Cambridge, MA, USA}

\author[0000-0002-4770-5388]{Ilaria Caiazzo}
\affiliation{Institute of Science and Technology Austria (ISTA), Am Campus 1, 3400 Klosterneuburg, Austria}

\author[0000-0002-2626-2872]{Jan van Roestel}
\affiliation{Anton Pannekoek Institute for Astronomy, University of Amsterdam, 1090 GE Amsterdam, The Netherlands}

\author[0000-0003-4373-7777]{Paula Szkody}
\affiliation{Department of Astronomy, University of Washington, 3910 15th Avenue NE, Seattle, WA 98195, USA}

\author{Andrew J. Drake}
\affiliation{Department of Astronomy, California Institute of Technology, 1200 East California Blvd, Pasadena, CA, 91125, USA}

\author[0000-0002-7004-9956]{David A. H. Buckley}
\affiliation{South African Astronomical Observatory, PO Box 9, Observatory 7935, Cape Town, South Africa}
\affiliation{Department of Astronomy, University of Cape Town, Private Bag, Rondebosch 7701, Cape Town, South Africa}
\affiliation{Department of Physics, University of the Free State, PO Box 339, Bloemfontein 9300, South Africa}

\author[0000-0002-5956-2249]{Stephen B. Potter}
\affiliation{South African Astronomical Observatory, PO Box 9, Observatory 7935, Cape Town, South Africa}
\affiliation{Department of Physics, University of Johannesburg, PO Box 524, Auckland Park 2006, South Africa}

\author[0000-0002-2761-3005]{Boris Gaensicke}
\affiliation{Department of Physics, University of Warwick, Coventry CV4 7AL, UK}

\author[0000-0002-9709-5389]{Kaya Mori}
\affiliation{Columbia Astrophysics Laboratory, Columbia University, New York, NY, USA}

\author[0000-0001-8018-5348]{Eric C. Bellm}
\affiliation{DIRAC Institute, Department of Astronomy, University of Washington, 3910 15th Avenue NE, Seattle, WA 98195, USA}

\author[0000-0001-5390-8563]{Shrinivas R. Kulkarni}
\affiliation{Department of Astronomy, California Institute of Technology, 1200 East California Blvd, Pasadena, CA, 91125, USA}

\author{Thomas A. Prince}
\affiliation{Division of Physics, Mathematics, and Astronomy, California Institute of Technology, 1200 East California Blvd, Pasadena, CA, 91125, USA}

\author[0000-0002-3168-0139]{Matthew Graham}
\affiliation{Department of Astronomy, California Institute of Technology, 1200 East California Blvd, Pasadena, CA, 91125, USA}

\author[0000-0002-5619-4938]{Mansi M. Kasliwal}
\affiliation{Department of Astronomy, California Institute of Technology, 1200 East California Blvd, Pasadena, CA, 91125, USA}

\author[0000-0003-4725-4481]{Sam Rose}
\affiliation{Department of Astronomy, California Institute of Technology, 1200 East California Blvd, Pasadena, CA, 91125, USA}

\author[0000-0003-4531-1745]{Yashvi Sharma}
\affiliation{Department of Astronomy, California Institute of Technology, 1200 East California Blvd, Pasadena, CA, 91125, USA}

\author[0000-0002-2184-6430]{Tomás Ahumada}
\affiliation{Department of Astronomy, California Institute of Technology, 1200 East California Blvd, Pasadena, CA, 91125, USA}

\author[0000-0003-3768-7515]{Shreya Anand}
\affiliation{Department of Physics, California Institute of Technology, 1200 East California Blvd, Pasadena, CA, 91125, USA}
\affiliation{Department of Physics, Stanford University, 382 Via Pueblo Mall, Stanford, CA 94305, USA}
\affiliation{Kavli Institute for Particle Astrophysics and Cosmology, P.O. Box 2450, Stanford University, Stanford, CA 94305, USA}

\author[0000-0001-9383-786X]{Akke Viitanen}
\affiliation{INAF-Osservatorio Astronomico di Roma, via Frascati 33, 00040 Monteporzio Catone, Italy}
\affiliation{Department of Physics and Helsinki Institute of Physics, Gustaf Hällströmin katu 2, 00014 University of Helsinki, Finland}

\author[0000-0002-9998-6732]{Avery Wold}
\affiliation{IPAC, California Institute of Technology, 1200 E. California
             Blvd, Pasadena, CA 91125, USA}

\author[0000-0001-9152-6224]{Tracy X. Chen}
\affiliation{IPAC, California Institute of Technology, 1200 E. California
             Blvd, Pasadena, CA 91125, USA}

\author{Reed Riddle}
\affiliation{Caltech Optical Observatories, California Institute of Technology, Pasadena, CA  91125}     

\author{Roger Smith}
\affiliation{Caltech Optical Observatories, California Institute of Technology, Pasadena, CA  91125}

\begin{abstract}
White dwarfs (WDs) are the most abundant compact objects, and recent surveys have suggested that over a third of WDs in accreting binaries host a strong (B $\gtrsim$ 1 MG) magnetic field. However, the origin and evolution of WD magnetism remain under debate. Two WD pulsars, AR Sco and J191213.72--441045.1 (J1912), have been found, which are non-accreting binaries hosting rapidly spinning (1.97-min and 5.30-min, respectively) magnetic WDs. The WD in AR Sco is slowing down on a  $P/\dot{P}\approx 5.6\times 10^6$ yr timescale. It is believed they will eventually become polars, accreting systems in which a magnetic WD (B $\approx 10-240$\;MG) accretes from a Roche lobe-filling donor, spinning in sync with the orbit ($\gtrsim 78$ min). Here, we present multiwavelength data and analysis of Gaia22ayj, which underwent an outburst in March 2022. We find that Gaia22ayj is a magnetic accreting WD that is rapidly spinning down ($P/\dot{P} = 6.1^{+0.3}_{-0.2}\times 10^6$ yr) like WD pulsars, but shows clear evidence of accretion, like polars. Strong linear polarization (40\%) is detected in Gaia22ayj; such high levels have only been seen in the WD pulsar AR~Sco and demonstrate the WD is magnetic. High speed photometry reveals a 9.36-min period accompanying a high amplitude ($\sim 2$ mag) modulation. We associate this with a WD spin or spin-orbit beat period, not an orbital period as was previously suggested. Fast (60-s) phase-resolved optical spectroscopy reveals a broad ``hump'', reminiscent of beamed cyclotron emission in polars, between 4000--8000\;\AA. We find an X-ray luminosity of $L_X = 2.7_{-0.8}^{+6.2}\times10^{32} \textrm{ erg s}^{-1}$ in the 0.3--8 keV energy range, while two VLA radio campaigns resulted in a non-detection with a $F_r < 15.8 \;\mu\textrm{Jy}$ 3$
\sigma$ upper limit. The shared properties of both WD pulsars and polars suggest that Gaia22ayj is a missing link between the two classes of magnetic WD binaries.
\end{abstract}

\section{Introduction}
Magnetic fields are ubiquitous in the Universe, from planetary scales such as that of the Earth to extragalactic scales of the intergalactic medium. In both cases, dynamos have been proposed to be the origin, and the dynamo theory has recently been invoked as the possible origin of strong magnetic fields in cooling white dwarfs (WDs) undergoing crystallization \citep[e.g.][]{2017isern, 2021schreiber, 2022ginzburg}.

It has been observed that over a third of accreting WDs host a magnetic WD \citep{pala2020, 2024rodriguez_erosita}. Such systems, known as magnetic cataclysmic variables (CVs), consist of a magnetic WD accreting from a Roche-lobe filling donor star, in the form of intermediate polars (IPs; $B\approx 1-10$ MG) or, more commonly, polars ($B\approx 10-240$ MG). Both polars and IPs channel accreted material through the WD magnetic poles onto the surface since, in both cases, the magnetospheric radius extends well past the surface of the WD. In polars, the WD spin is typically locked with the orbital period\footnote{Asynchronous polars are polar-like systems in which there is a $\lesssim10$\% difference between the WD spin and the orbital period \citep[e.g.][]{2023littlefield}, which in some systems has been explained by a nova outburst throwing the system out of synchronism \citep[e.g.][]{1991schmidth}.}, but in IPs, the WD spins $\sim10-100$ times faster than the orbital period. 

Despite the abundance of magnetic CVs, however, only $\sim 2$\% of their younger progenitors, detached post-common-envelope binaries (PCEBs), host a magnetic WD \citep[e.g.][]{2012pceb, 2021parsons}. Because CVs are further evolved and host cooler WDs, this has led to the idea that magnetism arises \textit{as a result} of a crystallization-driven dynamo as the WD cools \citep[e.g.][]{2021schreiber}. Other channels for WD magnetism, such as 1) the ``fossil field" scenario, where the progenitor was a magnetic Ap/Bp star \citep{2004Braithwaite}, 2) a dynamo operating during the common envelope phase (CE) \citep{2008tout}, and 3) a double WD merger \citep{2012garcia} are also possible, though the former two scenarios would predict a large number of detached PCEBs hosting a magnetic WD. In the dynamo scenario, a crystallizing WD accretes from a donor star, is spun up, and generates a $B\approx 1-250$ MG magnetic field \citep{2021schreiber}. The WD may then synchronize with the orbit, leading to the creation of a polar. Recent work has suggested that this idea may well lead to some, but not all of the polars hosting $\gtrsim10$ MG magnetic fields \citep{2022ginzburg, 2024camisassa}. 

Furthermore, two WD ``pulsars'' have been discovered in the last decade: AR Sco \citep{2016marsh} and J191213.72--441045.1 \citep[henceforth J1912;][]{2023pelisoli_j1912}, the first of which motivated a connection from the dynamo theory to reconcile the role of AR Sco in CV evolution \citep{2021schreiber}. Curiously, it appears that the WD temperature of J1912 is too high for crystallization to have taken place, potentially calling into question the dynamo origin of magnetic fields or simply showing that this channel may not lead to all magnetic CVs \citep{2024pelisoli}. Nevertheless, the evolutionary picture of \cite{2021schreiber} may still explain the origin of its magnetic field, as the recent work of \cite{2024camisassa} suggested that dynamo-generated magnetic fields can break out at higher temperatures for more massive WDs.

Regardless of the origin of magnetism in WD pulsars, these systems are close ($P_\textrm{orb}\approx 3.5 - 4$ h), \textit{detached, non-accreting} binaries, where fast, pulsed emission ($P_\textrm{spin} \approx 2-5$ min) out to radio frequencies has been detected. This emission has been attributed to the acceleration of particles in the interaction between the magnetic field ($B\approx 50-100$ MG) of the WD and that of the M dwarf companion, though whether the radio pulses are dominated by synchrotron or cyclotron emission is under debate \citep[e.g.][]{2016marsh, 2018stanway}. Crucially, both AR Sco \citep{2022pelisoli_arsco} and J1912 (Woudt, P. et al. in prep) have been observed to be rapidly spinning down, with AR Sco showing $P/\dot{P} = 5.6 \times10^6 \textrm{ yr}$. This suggests that these systems will eventually become polars---the donor star will fill its Roche lobe in a few Gyr due to gravitational wave radiation and magnetic braking , and the WD will spin down to synchronize with the orbital period.

 
Here, we report the characterization of Gaia22ayj as an accreting magnetic WD which will likely evolve into a polar. \cite{2022kato} first reported the discovery of Gaia22ayj after it underwent an optical outburst in March 2022, found a 9.36-min period in data from the Zwicky Transient Facility (ZTF) and proposed it to be a WD binary based on the claim that this was the orbital period. We present multiwavelength data to show that the 9.36-min period in Gaia22ayj \textit{is not} the orbital period of a binary system, rather it is likely the WD spin (or spin-orbit beat) period of an accreting magnetic WD. In this sense, Gaia22ayj is like an IP, but its large photometric amplitude and spectroscopic modulation are reminiscent of polars, therefore suggesting this is, at the very minimum, a new empirically-defined subclass of magnetic CVs. The extreme levels of optical linear polarization reaching 40\%, as presented in this work, are seen only in AR Sco \citep{2017buckley}, further hinting at Gaia22ayj being a possible connection between WD pulsars and polars.

Recently, WD pulsars have gained traction as possible explanations for ``long-period radio transients" (LPTs), which are radio sources pulsing on the timescales of minutes to hours \citep{2023hurleywalker}. In two systems, M dwarfs have been seen in optical spectra, though at the time of writing, spectroscopic confirmation of WDs in these systems remains to be obtained \citep{2024hurleywalker, 2024deruiter}. The possible connection of Gaia22ayj to WD pulsars that we present here brings to light the likely diversity that exists in WD pulsars and related systems. 

In Section \ref{sec:data}, we present optical photometry and spectroscopy, including polarimetry, that show that Gaia22ayj is an accreting magnetic WD in a close binary. In Section \ref{sec:multiwavelength_data}, we present all multiwavelength data collected on Gaia22ayj in the radio, infrared, X-ray, and (false) $\gamma$-ray association. In Section \ref{sec:analysis}, we show that the 9.36-min period likely represents the spin of the magnetic WD, and that it is rapidly slowing. In Section \ref{sec:discussion}, we argue for a possible interpretation of Gaia22ayj as a link between WD pulsars and polars. We also present estimates of the population of such objects and projections for the discovery of Gaia22ayj-like systems in the upcoming Rubin Observatory Legacy Survey of Space and Time (LSST).

\section{Optical Photometry and Spectroscopy}
\label{sec:data}

\subsection{Archival Photometry}
Gaia22ayj was first discovered by \cite{2022kato} using the \textit{Gaia} alerts stream. In Data Release 3 (DR3), Gaia22ayj has ID 5697000580270393088 and an associated distance of $2.5^{+1.5}_{-1.0}$ kpc as estimated by \cite{2021bailerjones} from its parallax of $\omega = 0.34\pm0.22$ mas. Upon querying the Zwicky Transient Facility (ZTF) photometric database, Gaia22ayj was shown to be periodic, at 9.36 min \citep{2022kato}. ZTF is a photometric survey which uses a 47 $\textrm{deg}^2$ field-of-view camera mounted on the Samuel Oschin 48-inch telescope at Palomar Observatory \citep{bellm2019, graham2019, dekanyztf, masci_ztf}. It uses custom $g$, $r$, and $i$ filters, taking most of its data in the $r$ filter at 30-s exposure times. Crucial to our characterization of Gaia22ayj was the following: in its first year of perations, ZTF carried out a nightly public Galactic Plane Survey in $g$ and $r$-band \citep{ztf_northernskysurvey_bellm} as well as a partnership survey which obtained
continuous (30 sec + 10 sec of read-out time) photometry of a Galactic Plane field for 1.5 hours, though some fields have up to 6 hours of continuous coverage \citep{ kupfer_ztf}. Since entering Phase II, the public Northern Sky Survey is at a 2-day cadence. The pixel size of the ZTF camera is 1$\arcsec$ and the median delivered image quality is 2.0$\arcsec$ at FWHM. 

We use ZTF forced photometry extracted at the position reported by \textit{Gaia} Data Release 3 \citep[DR3, and corrected to J2000;][]{2023gaia_dr3} of Gaia22ayj, including proprietary data to be made publicly available in upcoming data releases, taken through 01 Nov, 2024, processed by IPAC at Caltech\footnote{\url{https://irsa.ipac.caltech.edu/data/ZTF/docs/ztf_forced_photometry.pdf}}. This allows one to obtain flux estimates below the detection threshold as well as more realistic error bars on the data. 

We also query the forced photometry database of the Asteroid Terrestrial-impact Last Alert System \citep[ATLAS;][]{2018atlas} at the \textit{Gaia} position of Gaia22ayj in both ATLAS $c$ (cyan) and $o$ (orange) bands. Stringent quality cuts were applied to the data, ensuring $5\sigma$ flux measurements, a maximum sky brightness of 20.5 mag arcsec$^{-2}$, and a well-sampled PSF (3.5 pixels in each spatial direction). 

\begin{figure*}
    \centering
    \includegraphics[width=\textwidth]{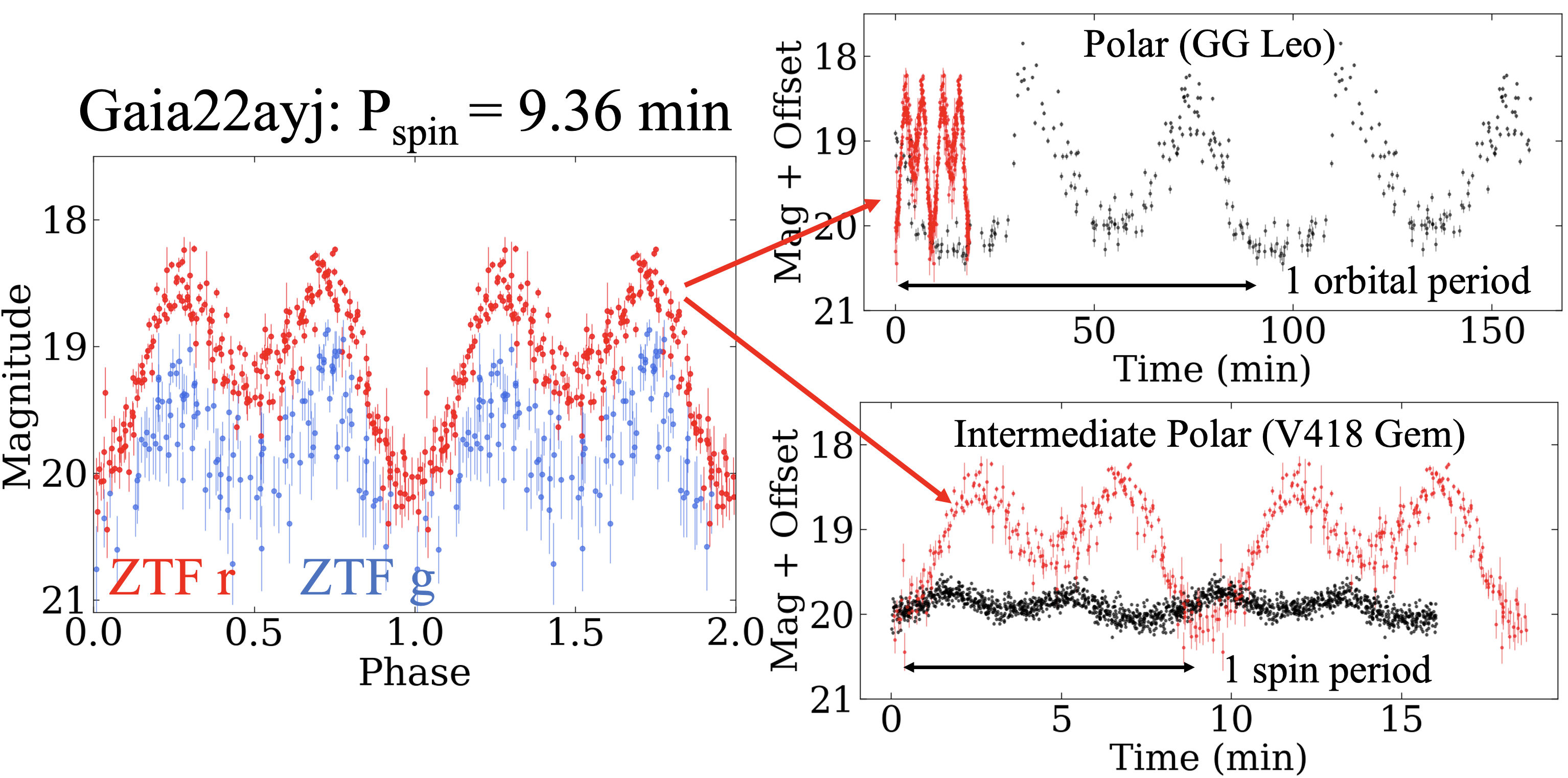}
    \caption{\textit{Left:} ZTF light curve of Gaia22ayj in $r$ and $g$ bands folded on the 9.36-min period. \textit{Right:} Comparison of the Gaia22ayj ZTF $r$ band light curve (red) to that of an archetypal polar, GG Leo (top; black points; $P_\textrm{spin}=P_\textrm{orb}=1.3$ h) and an archetypal IP, V418 Gem (bottom; black points; $P_\textrm{spin}= 8.0$ min). The light curves of GG Leo and V418 are offset to match with the minimum of Gaia22ayj. Gaia22ayj pulsates at the short period of an IP, but at the high amplitude of a polar.}
    \label{fig:main}
\end{figure*}

We performed a Lomb-Scargle period analysis \citep{1976lomb, 1982scargle} of ZTF data using \texttt{gatspy} \citep{gatspy}. The strongest peak is at 4.68 minutes, which we confirm as half of the period; folding the light curve on twice that period (9.36 min) reveals two different minima. We show the ZTF light curve of Gaia22ayj (excluding outbursting epochs) in Figure \ref{fig:main}, and compare the ZTF $r$ light curve to that of an archetypal polar, GG Leo, as well as to that of an archetypal IP, V418 Gem\footnote{V418 Gem is one of the IPs with the highest amplitude optical light curve, with some IPs showing little to no detectable variation in the optical on their spin periods (though in some cases they must be seen in the X-ray to confirm their IP nature).}. Both GG Leo and V418 Gem show double-peaked light curves like Gaia22ayj over a single WD spin period, as do many pre-polars \citep{2024vanroestel}. In brief, Gaia22ayj has the amplitude of a polar, but the rapid spin period of an IP. 

\subsection{Long Term Photometry}
\begin{figure*}
    \centering
    \includegraphics[width=\textwidth]{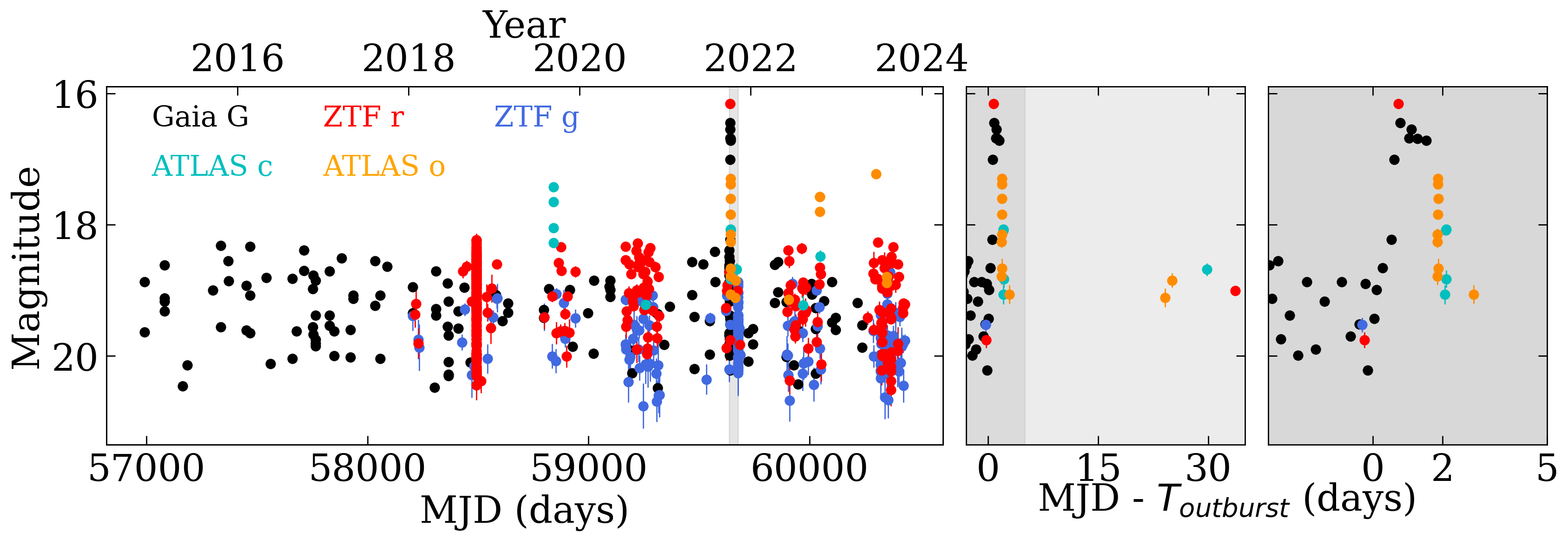}
    \caption{\textit{Gaia} coverage from 2014--2024 shows consistent high amplitude modulation, while both ZTF (2018--onwards) and \textit{Gaia} show a $\sim3$-mag outburst beginning on 3 April 2022. ATLAS coverage demonstrates that the outburst lasts two days, during which the high amplitude modulation seen in quiescence disappears. The low amplitude and short duration of the outburst more closely resembles those seen in IPs than those in non-magnetic dwarf novae. In either case, this outburst suggests ongoing accretion in Gaia22ayj.}
    \label{fig:long_term}
\end{figure*}

In Figure \ref{fig:long_term}, we show the long term behavior of Gaia22ayj in both \textit{Gaia}, ZTF photometry, and ATLAS photometry which spans ten years from 2014 to 2024. The ZTF photometric coverage is sparser than average (excluding deep drilling, $\sim50$ $r$-band epochs, whereas the ZTF average across the sky is $\sim850$ $r$-band points) due to the source's low declination. ATLAS photometry firmly establishes that the outburst lasted $\approx$2 days. In Figure \ref{fig:long_term}, \textit{Gaia} coverage shows that Gaia22ayj takes $\sim0.75$ day to rise to a peak brightness of $\sim16$ mag (starting from $\sim19$ mag), at which point the light curve is roughly constant for $\sim1$ day. ATLAS $c$ and $o$ band photometry shows a rapid decline back to $\sim19$ mag two days after outburst. Approximately 30 days after the outburst, the ZTF and ATLAS data show Gaia22ayj still in quiescence.  These outbursts are on the lower amplitude end of dwarf nova outbursts, which typically range between 2--4 mag \citep[though low-accretion rate WZ Sge-like systems can reach 8 mag outbursts; e.g.][]{1995warner}. 

Instead of typical dwarf nova outbursts, the 2022 outburst of Gaia22ayj more closely resembles very short outbursts seen in IPs, notably in the famous systems V1223 Sgr \citep{2017hameury, 2022hameury} and TV Col \citep{1993hellier}. However, those outbursts last less than a day and do not show rapid fading like that of Gaia22ayj. \cite{2022scaringi} recently suggested that such bursts, increases in brightness by a factor of three over a few hours, are ``micronovae" which are localized thermonuclear events associated with magnetic WDs. It is unclear if the 2022 outburst of Gaia22ayj is related to such phenomena. During outburst, Gaia22ayj does not show the characteristic high amplitude modulation it does in quiescence, or the modulation seen in other IPs during outburst, though the \textit{Gaia} temporal cadence may not be high enough to reveal fast variability. There is some evidence of additional outburst-like behavior from the ATLAS light curve in 2019 and 2022, but aside from the outburst of 3 March 2022, no other such events were covered by multiple photometric surveys.

\begin{figure}
    \centering
    \includegraphics[width=0.45\textwidth]{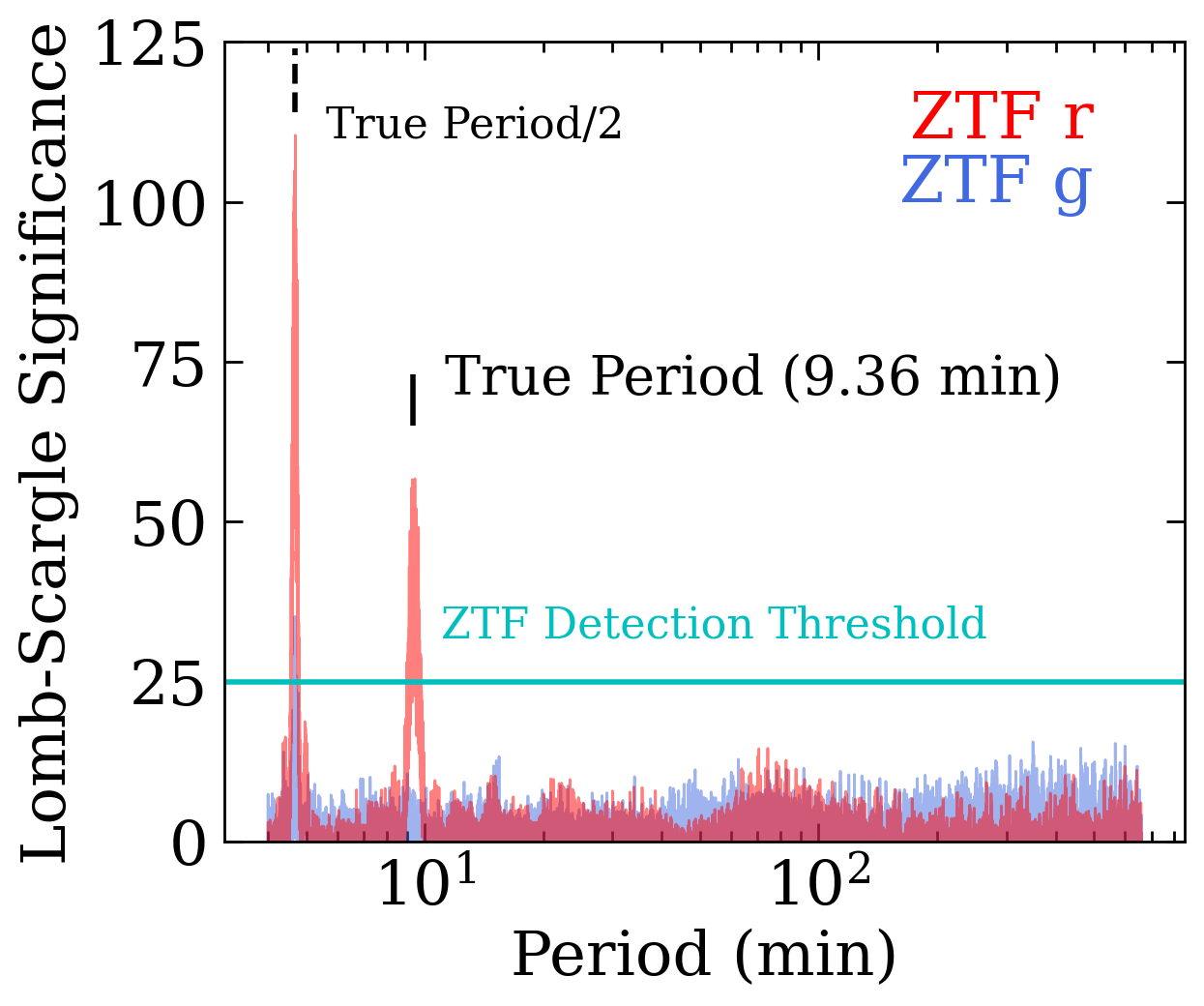}
    \caption{A Lomb-Scargle periodogram constructed from ZTF $r$ and $g$ photometry only reveals peaks at 9.36 min (true period) and 4.68 min (half of that). No other peaks, including one corresponding to a possible orbital period, pass the typical ZTF detection threshold of 25 in these units (see text for details).}
    \label{fig:periodogram}
\end{figure}

Finally, in Figure \ref{fig:periodogram}, we show the Lomb-Scargle periodogram constructed from the ZTF $r$ and $g$ light curves, showing clear peaks at 9.36 min (true period) and 4.68 min (half of that).  In Figure \ref{fig:periodogram}, we calculate a Lomb-Scargle periodogram using \texttt{gatspy} \citep{gatspy}, evenly sampling fifty thousand trial periods in frequency space between four minutes and eleven hours (in order to avoid strong harmonics of the sidereal day around twelve hours). We define the significance as the number of median absolute deviations above the median power. In samples of $\approx$10,000 ZTF light curves, we have found that a value of 25 in these units corresponds to approximately the 95$^\textrm{th}$ quantile. This means only five percent of ZTF light curves have a significance this high, which typically contain a true periodic signal. There are no significant peaks aside from the 4.68 and 9.36 min ones that pass this threshold, suggesting that long term photometry alone is unsuitable for detecting an orbital period.

\subsubsection{High Speed Photometry}
We obtained high speed photometry of Gaia22ayj on seven separate occasions: 25 April 2022 with ULTRACAM \citep{ultracam} on the 3.58-m New Technology Telescope at La Silla ($u, g, i$ filters simultaneously at 10-s cadence), 18 May 2023 with the Sutherland High Speed Optical Cameras (SHOC) on the 1m SAAO telescope (clear filter), 13 November 2023 and 6 January 2024 with the Caltech HIgh-speed Multi-color camERA \citep[CHIMERA;][]{chimera} on the 5-m Hale Telescope at Palomar Observatory ($g, r$ and $g, i$ filters simultaneously at 10-s cadence), 6 March 2024 and 16 April 2024 with HiPERCAM \citep{hipercam} on the 10.4-m Gran Telescopio Canarias at the Observatorio del Roque de los Muchachos on La Palma ($u,g,r,i,z$ filters simultaneously at 3.77-s cadence). All data were acquired along with GPS timestamps to ensure sub-millisecond timing precision, and corrected to a barycentric Julian date in barycentric dynamical time (BJD$_\textrm{TDB}$). All data were extracted using aperture photometry pipelines\footnote{\url{https://github.com/HiPERCAM/hipercam}} which computed the flux relative to the same star, Gaia DR3 5697012365660670720.  

\begin{figure*}
    \centering
    \includegraphics[width=0.5\textwidth]{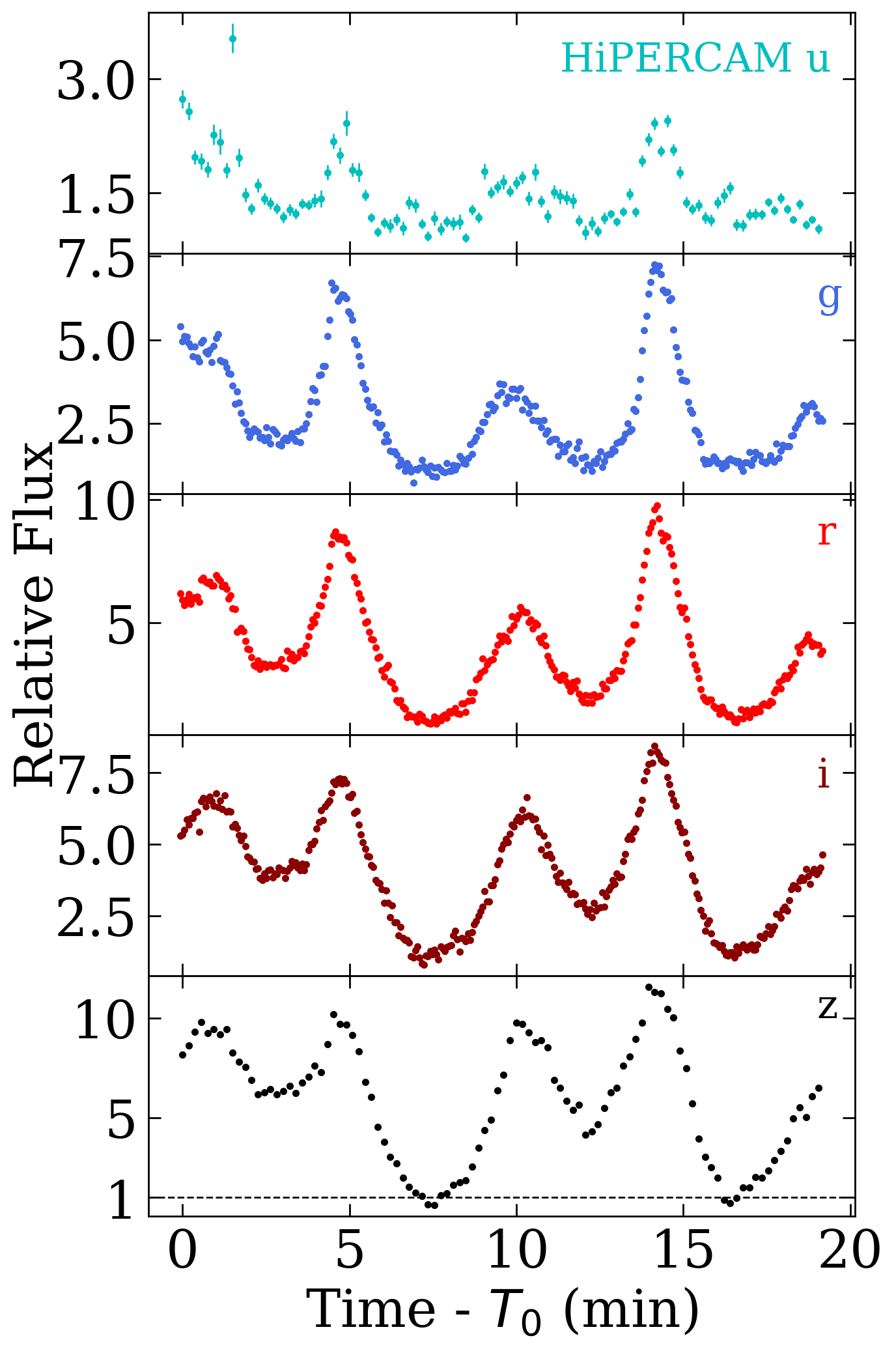}\includegraphics[width=0.5\textwidth]{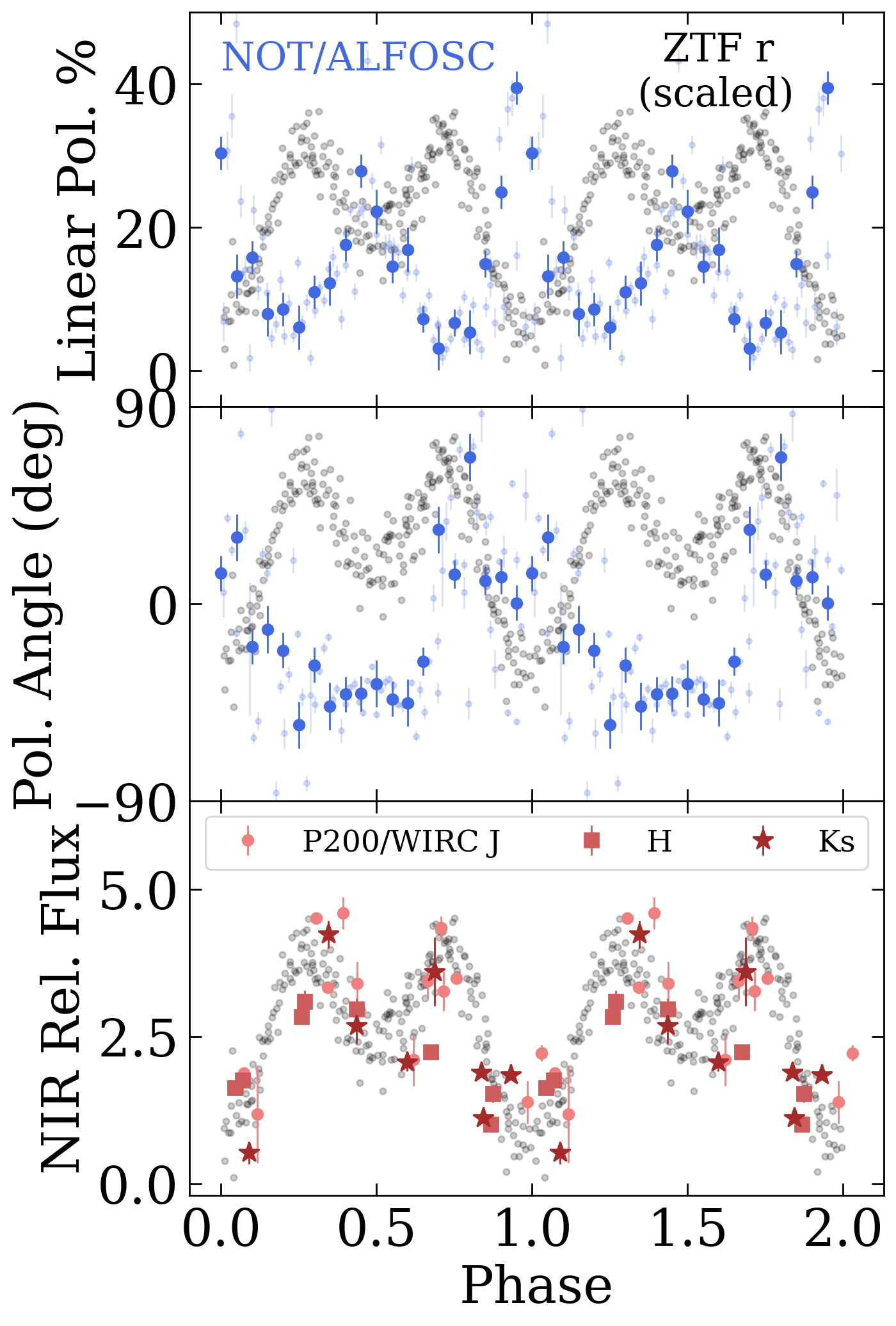}
    \caption{\textit{Left:} Quintuple-band simultaneous high speed (3.77 s) photometry of Gaia22ayj acquired over 20 min with HiPERCAM on the GTC shows that Gaia22ayj can increase in brightness by a factor of $\sim 10$ in 2.5 minutes, and that the variability amplitude varies significantly with wavelength, being lowest in the u band. Gaia22ayj also shows high levels of linear polarization (\textit{upper right}), with two peaks ($\sim20$ \% and $\sim40$ \%) anticorrelated with the peaks of the ZTF light curve (black, arbitrarily scaled). Such high levels of linear polarization are only rivaled by AR Sco. The double-peaked nature of the linear polarization curve, along with the polarization angle swing (\textit{middle right}), may suggest two-pole accretion. Triple-band near-infrared photometry (\textit{lower right}) reveals similar extreme behavior as in the optical (black; arbitrarily scaled).}
    \label{fig:hipercam}
\end{figure*}

In Figure \ref{fig:hipercam}, we show the 5-band simultaneous light curve acquired by HiPERCAM on 16 April 2024. Two maxima per spin period are clearly seen, as in the ZTF light curve, but the multi-band coverage shows that one peak is higher than the other at bluer bands. We show that the amplitude of the higher peak is variable in the full optical high speed light curves taken with P200/CHIMERA and NTT/ULTRACAM in Appendix \ref{sec:appendix}.

\subsection{Optical Polarimetry}
Linear optical polarimetry was taken with the Alhambra Faint Object Spectrograph and Camera (ALFOSC) on the 2.56-m Nordic Optical Telescope (NOT), La Palma, on three occasions: 1 May 2022, 2 May 2022, and 13 April 2024. ALFOSC was equipped with WeDoWO, a four-beam polarimetric unit, capable of obtaining a full linear polarimetric observation from a single exposure using a wedged double Wollaston prism. A red pass (OG570) filter was used to select the red part of the spectrum, where the photometric modulation is strongest. This setup produced 
a bandpass ranging from 5500\AA\ to $\sim$10000\AA, where the CCD response effectively tapers off. On all occasions, a comparison star was simultaneously monitored which showed no $>1\sigma$ changes in linear polarization percentage ($<1\%$) or angle. In Figure \ref{fig:hipercam}, we show the linear polarization (percentage and angle) of Gaia22ayj folded on the spin period. Data are taken from the third observing run, which lasted one hour and took place in excellent conditions. The entire dataset is folded on the 9.36-minute period (light blue) and averaged over twenty phase bins (dark blue). Two distinct peaks are seen, with one reaching 
$\sim20$ \% polarization and the other $\sim40$ \% polarization, each coincident with the two minima in total flux. The same behavior is seen two years apart, in the 2022 as well as the 2024 observations of Gaia22ayj. The full light polarimetric light curve is shown in the Appendix Figure \ref{fig:lin_pol_tot}.

\subsection{Spectroscopy}
We obtained optical spectroscopy of Gaia22ayj on various occasions. The first spectrum was taken with the Low Resolution Imaging Spectrometer \citep[LRIS;][]{lris} on the Keck I telescope on 3 April 2022 through some cloud coverage, but showed emission lines, ruling out a detached WD binary nature. The next set of spectra, acquired on 10 April 2022 with the Magellan Echellete Spectrograph \citep[MagE;][]{mage} on the Magellan Clay Telescope in April 2022 confirmed that emission lines were present in Gaia22ayj and showed that an overall modulation across the entire optical band was responsible for the high amplitude photometric modulation. 

Most notably, we acquired low-resolution ($\Delta \lambda \approx 0.8$ \AA) spectra with Keck I/LRIS on 8 November 2023 (for a total time of 1.6 h) and on 19 November 2023 (for a total time of 1.96 h) at high speed 60-s exposures on the blue channel, 90 s on the red with 2x2 and 2x1 binning (spatial vs spectral), respectively to sample the 9.36-min spin period. The first run had strong winds and some light clouds with 0.9" seeing, while the second run had more favorable conditions with 0.8" seeing that led to higher signal-to-noise ratio. All LRIS data were wavelength calibrated with internal lamps, flat fielded, and cleaned for cosmic rays using \texttt{lpipe}, a pipeline for LRIS optimized for long slit spectroscopy \citep{2019perley_lpipe}. 

\begin{figure*}
    \centering
    \includegraphics[width=0.5\textwidth]{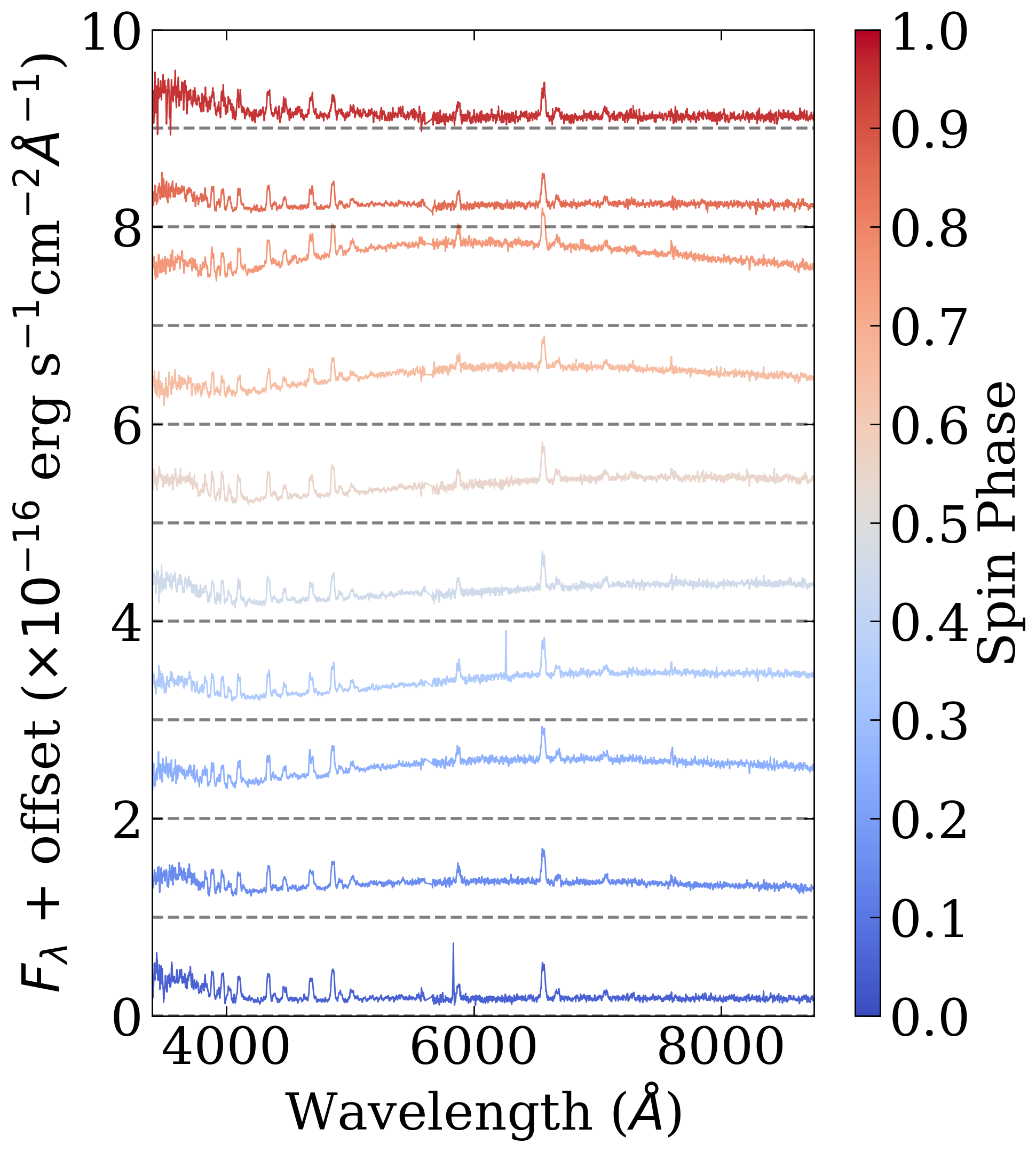}
    \includegraphics[width=0.2\textwidth]{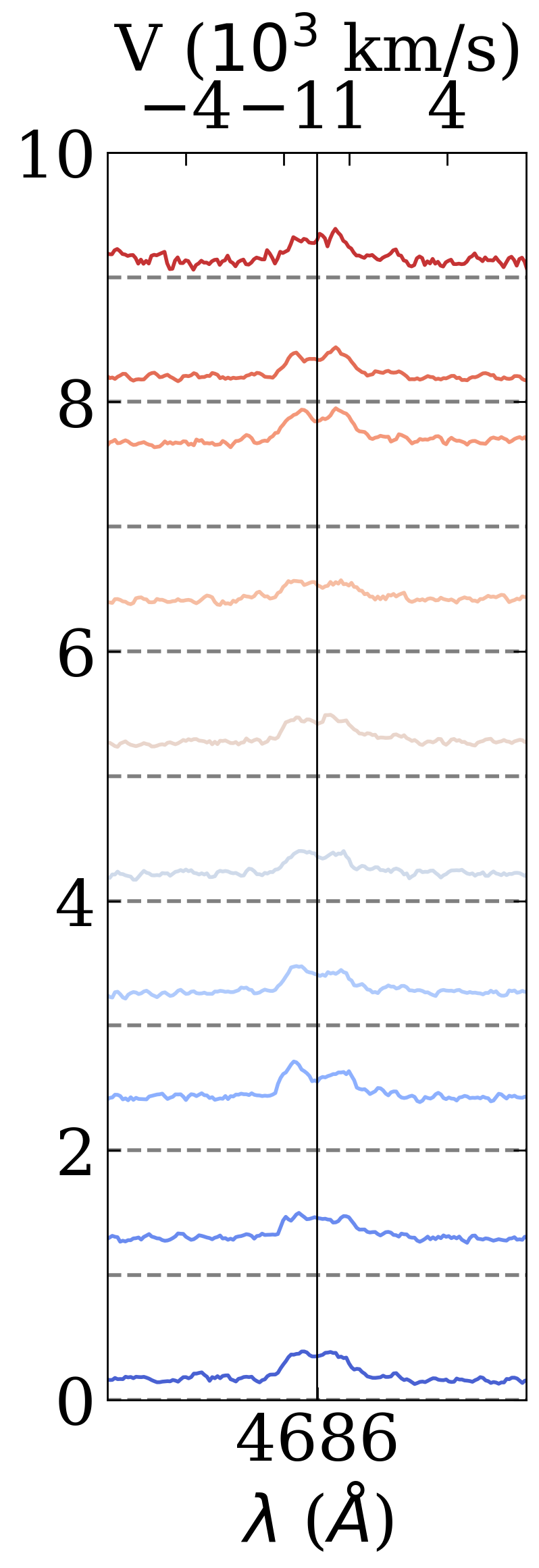}
    \includegraphics[width=0.2\textwidth]{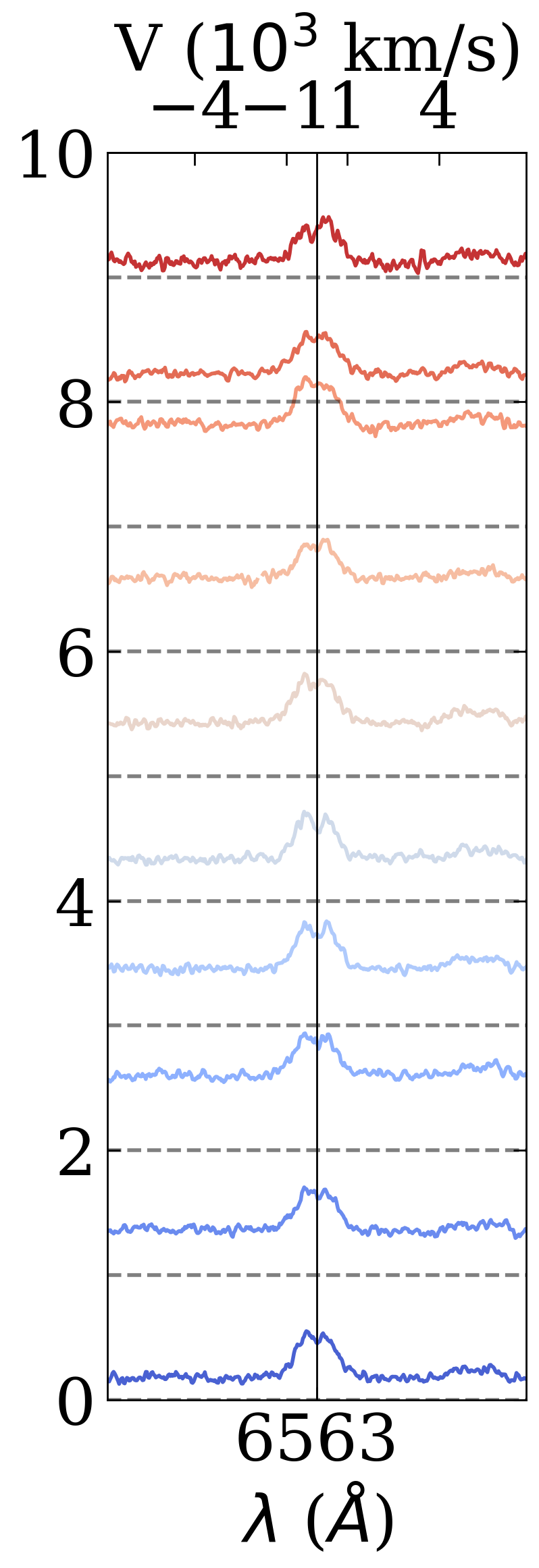}\\
    \includegraphics[width=\textwidth]{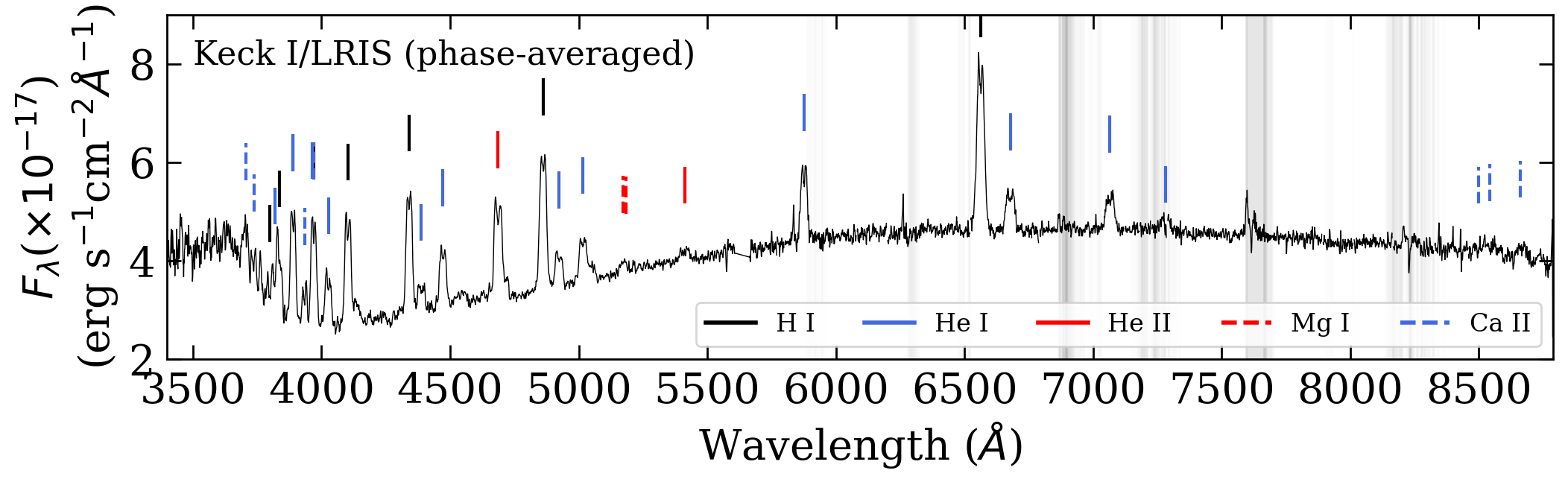}     
    \caption{Phase-resolved spectroscopy of Gaia22ayj shows that overall modulation between 4000--8000\AA\ leads to the observed high-amplitude photometric variability. Gray shaded areas are telluric features. \textit{Upper left}: Two hours of Keck I/LRIS spectra, stacked into ten bins folded on the 9.36-min period show two distinct maxima that resemble ``cyclotron humps", peaking at phases 0.25 and 0.75 (third and eighth sub-panels from the bottom, respectively). \textit{Upper right}: The He\,{\sc ii} 4686 and H$\alpha$ emission lines remain flat or double-peaked (broadened with $v\approx1200~\textrm{km s}^{-1}$) and show no amplitude or RV modulation on the spin phase. \textit{Bottom}: The phase-averaged, stacked spectrum of Gaia22ayj reveals prominent H and He (slightly double-peaked) emission lines and a Balmer jump in emission, commonly seen in CVs. The high concentration of hydrogen rules out an ultracompact nature. Strong He\,{\sc ii} 4686 (He II/H$\beta$ $\approx$ 1) is suggestive of a magnetic WD. Ca II and Mg I emission lines are marginally detected, which could trace the irradiated face of the donor star.}
    \label{fig:spec}
\end{figure*}

In Figure \ref{fig:spec}, we show phase-resolved and averaged spectra from the second run, which confirm the initial MagE findings, though with higher temporal and spectral resolution. The slightly double-peaked nature of the Balmer, He I, and He\,{\sc ii} emission lines suggests either the presence of a disk or face-on viewing of two accretion poles. Metal lines in emission (Mg I and Ca II) are seen only in the phase-averaged spectrum, though even then with just a marginal detection. If indeed present, these lines likely trace the irradiated face of the donor star \citep[e.g.][]{2023rodriguez}. The Na I doublet at 8183 and 8195 \AA\; in late-type stars is not seen.

\begin{figure}
    \centering
    \includegraphics[width=0.45\textwidth]{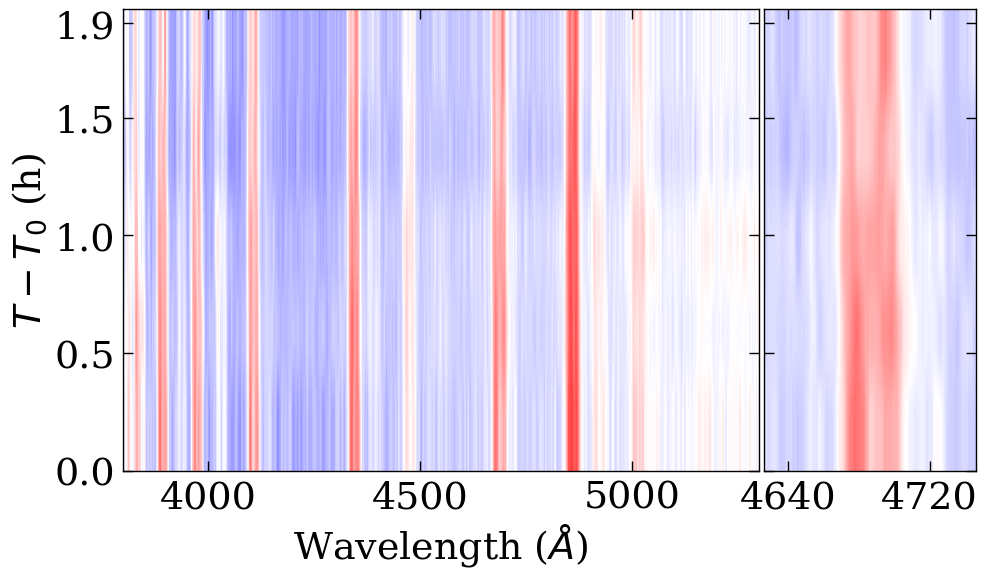}\\
    \includegraphics[width=0.45\textwidth]{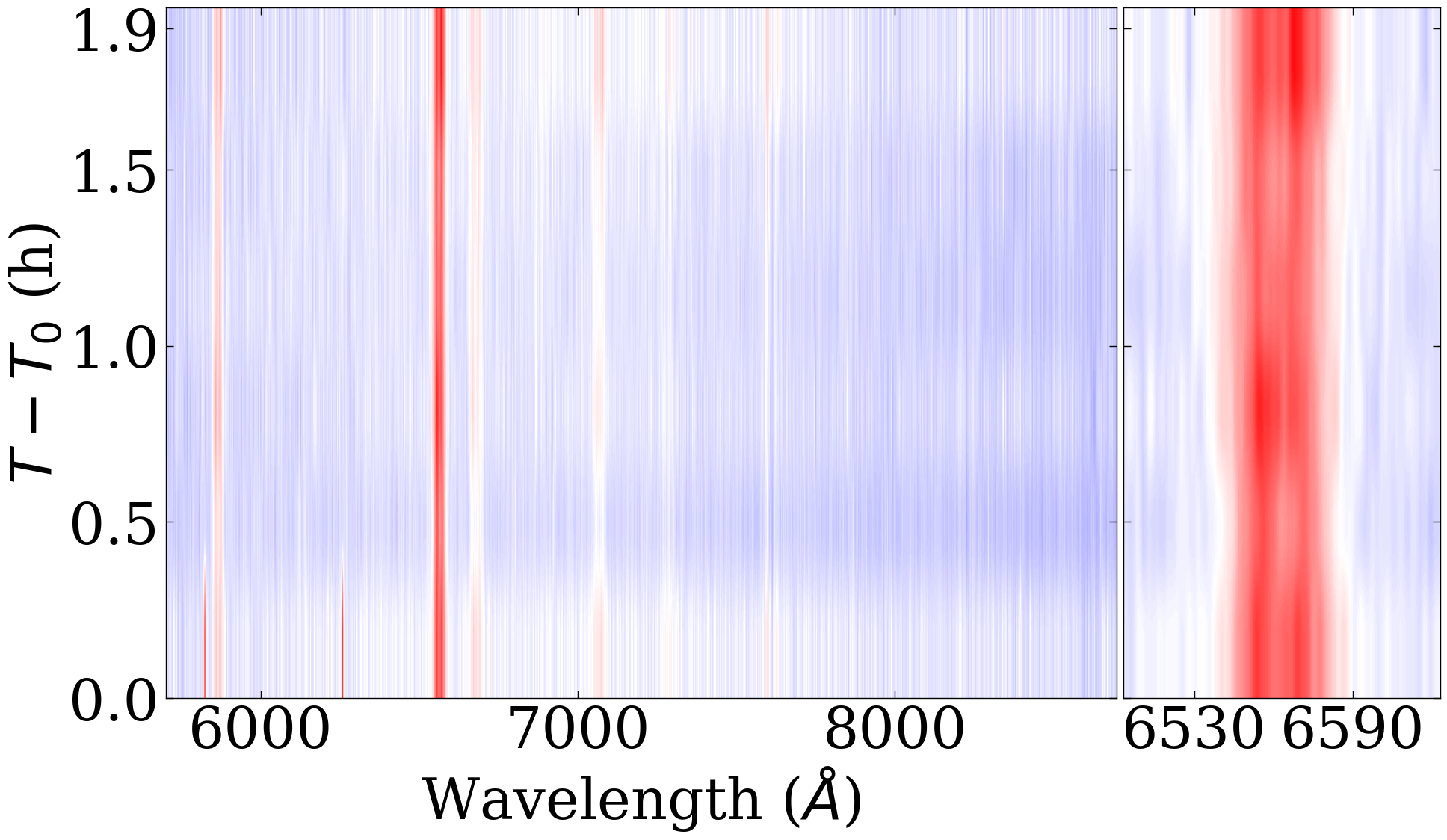}
    \caption{Trailed Keck I/LRIS continuum-normalized spectra acquired over $\approx 2$ h do not reveal any RV shifts in emission lines down to the limiting resolution of $\approx 30$ km s$^{-1}$.}
    \label{fig:total_spec}
\end{figure}

In Figure \ref{fig:total_spec}, we show the trailed Keck I/LRIS spectra of the second run, acquired over 1.96 h. No radial velocity (RV) shifts of the Balmer, He I, or He\,{\sc ii} emission lines or suspected donor lines (Mg I and Ca II) are seen. We experimented by binning a different number of individual spectra and using different smoothing levels to probe RV shifts, but none are seen down to the limiting resolution of our LRIS setup ($0.8~\mathrm{\AA} = 30~\textrm{km s}^{-1}$). 

\subsubsection{Spectropolarimetry}
\begin{figure}
    \centering
    \includegraphics[width=0.5\textwidth]{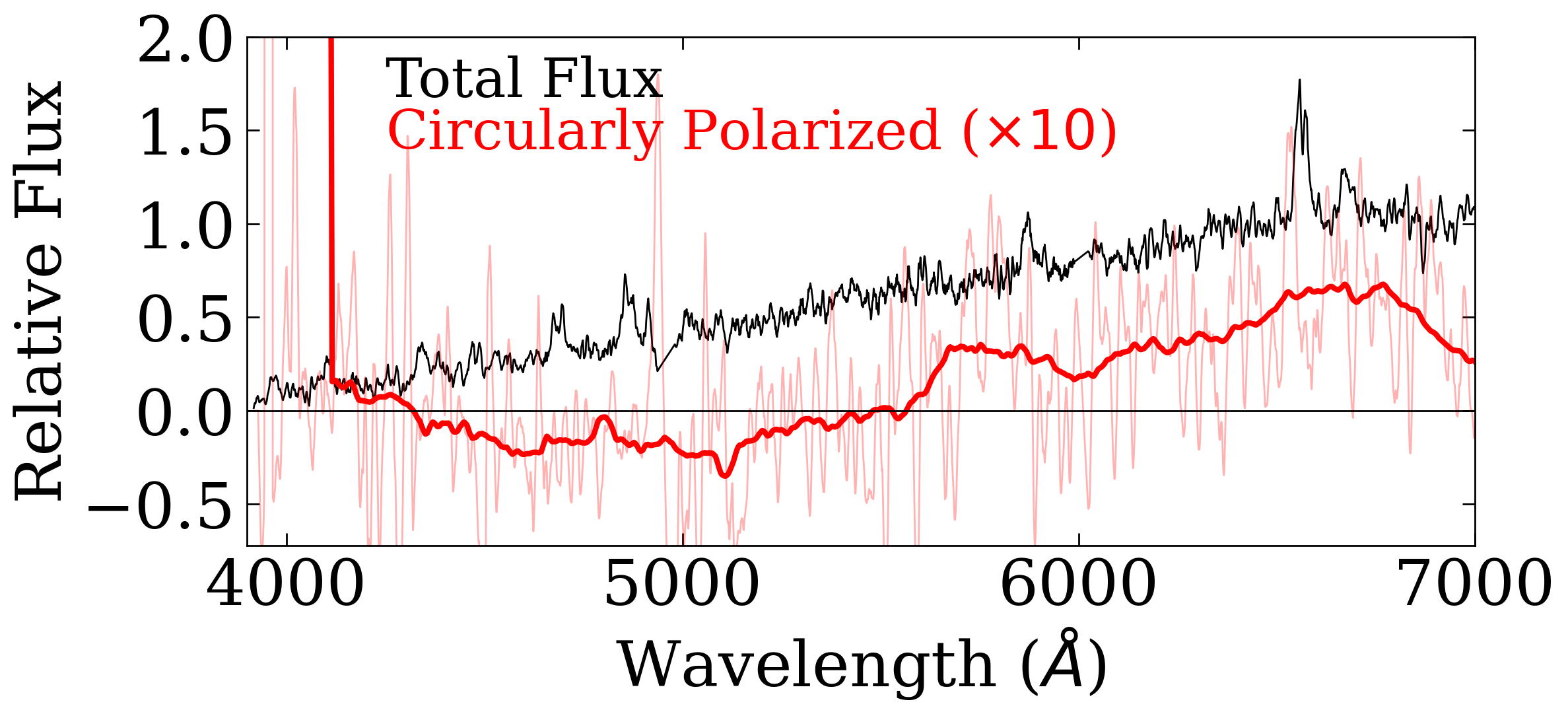}
    \caption{Low-resolution spectropolarimetry acquired with SALT reveals a possible ($>1\sigma$) detection of a circularly polarized continuum, peaking around 6800 \AA. The height of the feature corresponds to a five percent level of circular polarization, consistent with magnetic CVs. }
    \label{fig:spectropol}
\end{figure}

We obtained spectropolarimetry of Gaia22ayj on 28 April 2022 using the Robert Stobie Spectrograph (RSS; \citealp{Burgh_etal_2003, Kobulnicky_etal_2003}) in circular spectropolarimetry mode \citep{spectropol} on the 10-m Southern African Large Telescope (SALT; \citealp{Buckley_etal_2006}). The observations consisted of two exposures of 600 sec for each of two positions of a 1/4 waveplate retarder, which was repeated once. In Figure \ref{fig:spectropol}, we show the resulting data: total flux (black) and  circularly polarized spectrum (original in light red; smoothed in red). We detect circularly polarized flux, reaching up to $\sim$5\% level at $\sim$6800\AA, with a $>1\sigma$ significance.  A nearby comparison star on the slit showed polarisation consistent with zero percent. Due to the long exposure times, these observations are unable to resolve the 9.3 min spin period. Further data are needed to confirm this detection and search for variability on longer time scales.

\section{Multiwavelength Observations}
\label{sec:multiwavelength_data}

\subsection{Near-Infrared Photometry}
We observed Gaia22ayj on 7 November 2023 in $J$, $H$, and $K_s$ bands with the Wide field InfraRed Camera (WIRC) on the 200-inch Hale Telescope at Palomar Observatory. A 9-dither observing pattern, tiling the field by shifting 10$\arcsec$ in a 3x3 square, was used. We acquired single 45-s $J$-band exposures, co-added six 9-s $H$-band exposures, and co-added ten 3-s $K_s$-band exposures, with exposures taken in each filter for a total of forty minutes. Data were dark-subtracted and flat fielded using standard techniques implemented in a custom \texttt{wirc\_pipe} pipeline (De, K. and Karambelkar, V. in prep.).

In Figure \ref{fig:hipercam}, we show the WIRC light curves folded on the 9.36-min optical period. Because we had to bin multiple exposures, we could not independently solve for a period using the near infrared data alone. All light curves appear to have a similar behavior as in the optical, peaking twice per spin period.

\subsection{X-ray Detection}

\begin{table}
\fontsize{8}{12}\selectfont
\renewcommand{\arraystretch}{0.75}
\centering
\caption{The best-fit spectral parameters and their errors for the different models applied to analyze the combined Swift/XRT X-ray spectrum of Gaia22ayj. }
\label{tab:xfit}
\begin{tabular}{lcc}
\hline                     

\multicolumn{2}{c} { Model: ${\tt tbabs\times powerlaw }$ }           \\
\hline
{\it Parameters:}               &                                \\
 $N_{\rm H}$ ($\rm \times 10^{22}\ cm^{-2}$) & $\la 0.09$ \\
$\Gamma$       &  $\rm 1.27^{+0.15}_{-0.14}$    \\
 $\rm \chi^2_{red}$ (dof)   & $\rm 0.99(62)$   \\ 
\hline     

\multicolumn{2}{c} { Model: ${\tt tbabs\times mekal }$ }           \\
\hline
{\it Parameters:}                  &                                \\
 $N_{\rm H}$ ($\rm  \times 10^{22}\ cm^{-2}$) &  $\rm 0.03^{+0.04}_{-0.03}$ \\
 ${\rm k}T$ (keV)    &  $\ga 15$    \\
$\rm \chi^2_{red}$ (dof) &  $\rm 0.98(62)$   \\ 
\hline

\multicolumn{2}{c} { Model: ${\tt tbabs\times mkcflow }$ }           \\
\hline
{\it Parameters:}                  &                                \\
$N_{\rm H}$ ($\rm \times 10^{22} \ cm^{-2}$) &  $\rm 0.07^{+0.05}_{-0.04}$ \\
 ${\rm k}T_{\rm min}$ (keV)     &  $\rm 0.1\ (fixed)$   \\
 ${\rm k}T_{\rm max}$ (keV)     &  $\ga 64$   \\
 $\dot{M}\ (\times 10^{-12}\ M_{\odot}\, \rm yr^{-1})$  &  $\rm 28.2^{+5.1}_{-2.4}$    \\
$\rm \chi^2_{red}$ (dof)  & $\rm 1.12(62)$   \\ 
\hline

\multicolumn{2}{c} { Model: ${\tt tbabs\times pcfabs \times mkcflow }$ }   \\
\hline
$N_{\rm H}$ ($\rm \times 10^{22}\ cm^{-2}$) & $\rm 0.02^{+0.05}_{-0.02}$ \\
$N_{\rm H,pc}$ ($\rm \times 10^{22}\ cm^{-2}$) &   $\rm 1.96^{+4.10}_{-1.36}$ \\
$\rm pcf$ (per cent) & $\rm 38^{+32}_{-20}$ \\
 ${\rm k}T_{\rm min}$ (keV)     &  $\rm 0.1\ (fixed)$   \\
${\rm k}T_{\rm max}$ (keV)       &  $\rm \ga 13$ \\
$\dot{M}\ (\times 10^{-12}\ M_{\odot}\, \rm yr^{-1})$  &  $\rm 38.1^{+130.2}_{-8.4}$    \\
 $\rm \chi^2_{red}$ (dof)   &  $\rm 1.01(60)$   \\ 

\hline
\hline
 $F_{\rm obs}$ ($\rm \times 10^{-13}\ erg\ s^{-1} cm^{-2}$) & $\rm 3.3\pm0.3$\\
$F_{\rm un}$ ($\rm \times 10^{-13}\ erg\ s^{-1} cm^{-2}$) & $\rm 3.9\pm0.9$\\
 $F_{\rm bol}$ ($\rm \times 10^{-13}\ erg\ s^{-1} cm^{-2}$) & $\rm 5.3\pm1.0$\\

\hline
$L_{\rm un}$ ($\rm \times 10^{32}\ erg\ s^{-1}$) & $\rm 2.7_{-1.8}^{+4.3}$\\
$L_{\rm bol}$ ($\rm \times 10^{32}\ erg\ s^{-1}$) & $\rm 3.7_{-2.4}^{+5.8}$\\
\hline
\end{tabular}
\flushleft
Notes: Upper and lower limits are computed for a 90\% confidence level. Observed ($F_{\rm obs}$) and absorption-corrected ($F_{\rm un}$) fluxes in the 0.3--8 keV energy band. Bolometric flux ($F_{\rm bol}$) in the 0.001--100 keV energy band. $F_{\rm un}$ and $F_{\rm bol}$ are computed for the {\tt mkcflow} component of the  $\tt tbabs\times pcfabs \times mkcflow$ model.
\end{table}

\begin{figure*}
\centering
\includegraphics[width=0.48\textwidth,clip=true]{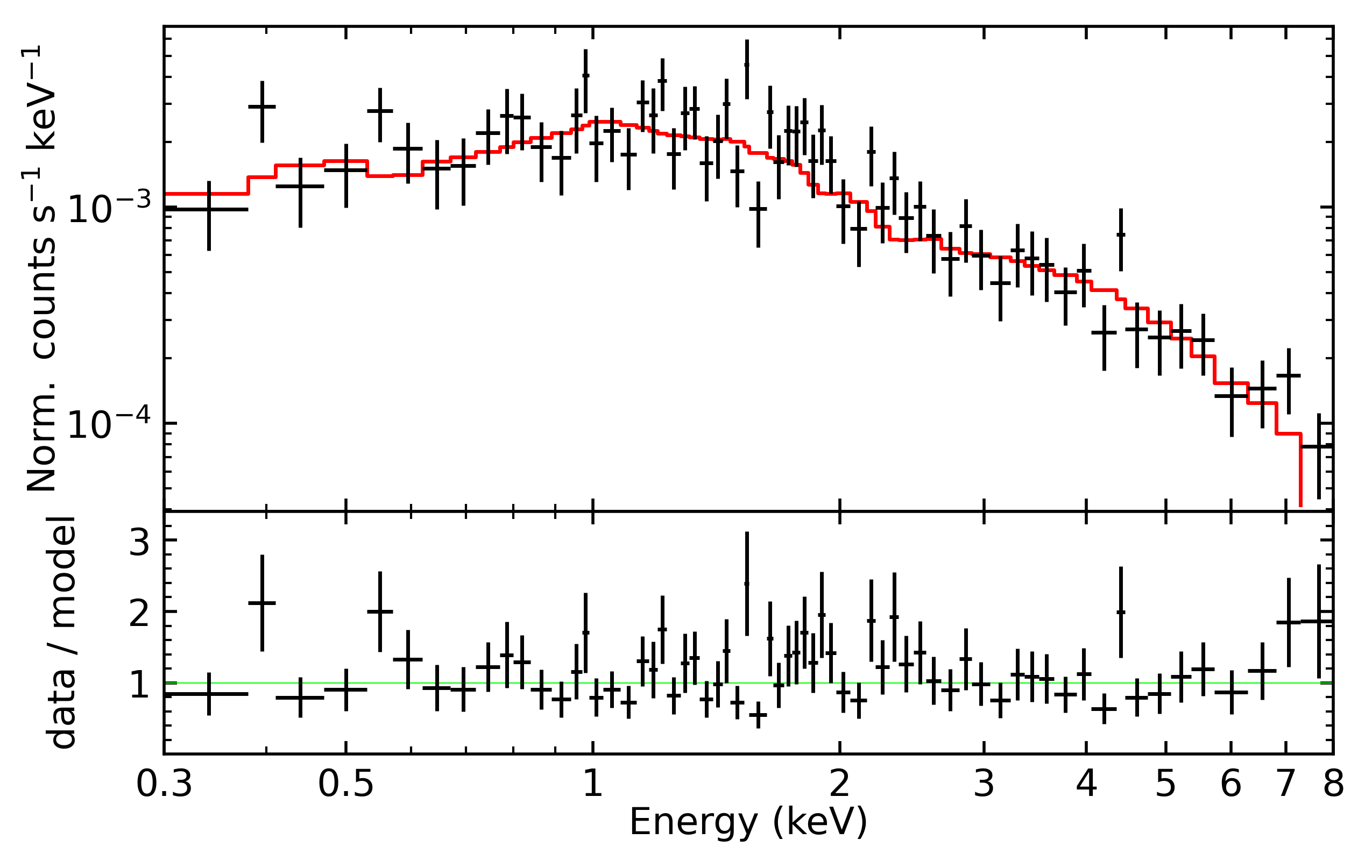}\includegraphics[width=0.44\textwidth,clip=true]{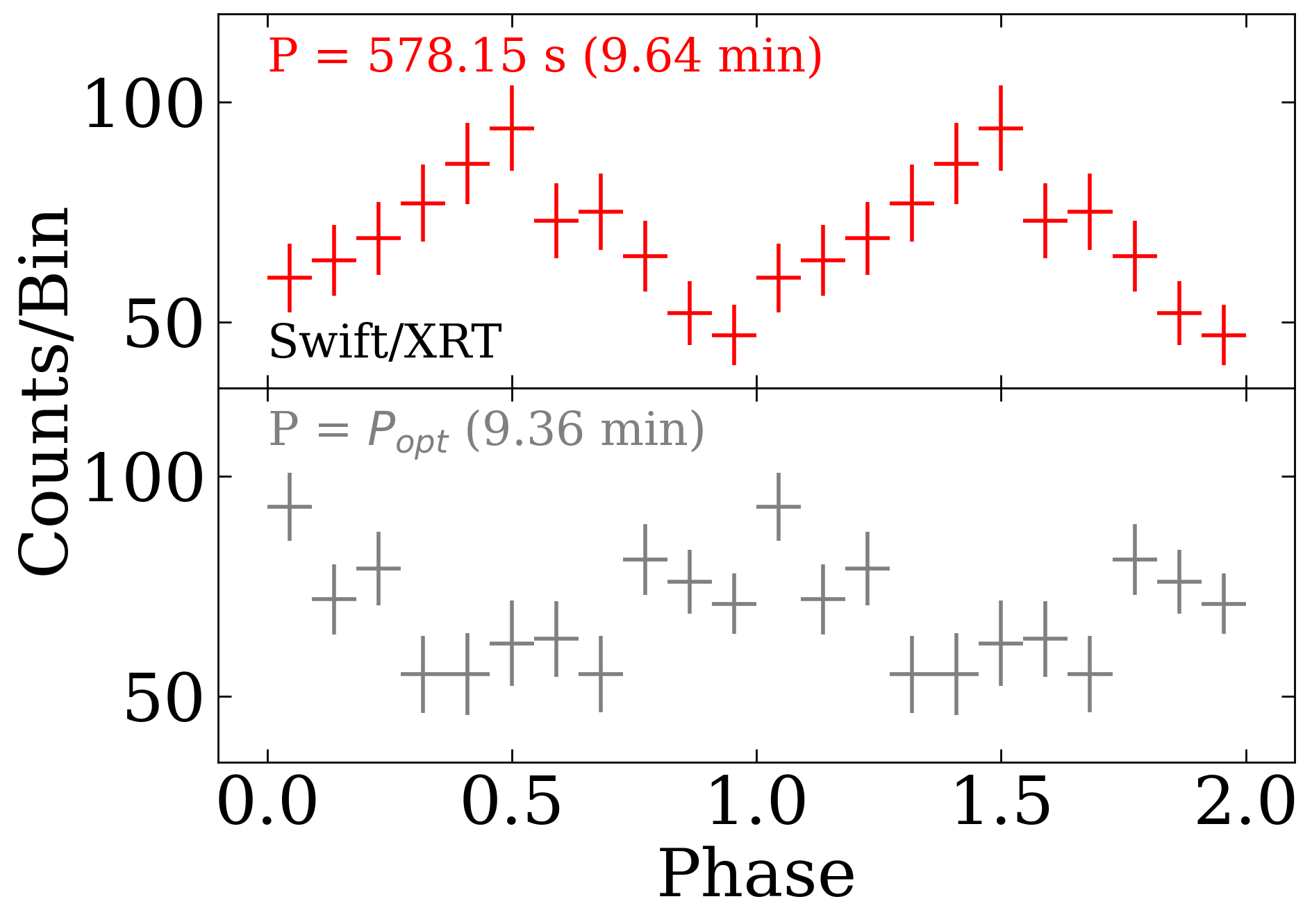}
\caption{\textit{Left: }The combined X-ray spectrum of Gaia22ayj from all 11 Swift/XRT observations (from June 21, 2005 to  December 6, 2005). {\it The red line} shows the best-fit ${\tt tbabs\times pcfabs \times mkcflow }$ model. {\it The bottom panel} shows the ratio of the data divided by the model spectrum. \textit{Right: } The X-ray light curve folded on the X-ray derived period (\textit{top}) and on the optically derived period (\textit{bottom}) is shown. Due to overlap with harmonics of the \textit{Swift} good time interval (GTI) of the observation, the X-ray period of 9.64 min is tentative and should be tested with further observations.}
\label{fig:Xray_spectrum}
\end{figure*}

Gaia22ayj is listed in the Second Swift-XRT Point Source Catalog (2SXPS) as 2SXPS J082526.4--223212 \citep{2020ApJS..247...54E}. The angular separation between the ZTF optical and Swift X-ray centroids is 1.2$\arcsec$, and the Swift X-ray 90$\%$ error circle is 2.2$\arcsec$, strongly suggesting this as the true X-ray counterpart. Swift/XRT serendipitously observed this source for a total of 86-ks exposure (at the location of the source on the detector), separated into 11 different observations.  This was due to a bright and well-studied BL Lac, QSO B0823--223, 8.25$\arcmin$ away, which was the target of the Swift/XRT observations. 

\subsubsection{X-ray Spectral Analysis}

We used the online web tool\footnote{\url{https://www.swift.ac.uk/user\_objects/}} to build Swift-XRT products \citep{2007A&A...476.1401G,2009MNRAS.397.1177E,2020ApJS..247...54E} and extract a single combined spectrum from all 11 observations of Swift/XRT (PC mode). To analyze X-ray spectra, we used the XSPEC v12.13 spectral modeling package \citep{1996ASPC..101...17A}.  We grouped spectral channels using the {\tt grppha} tool from FTOOLS \citep{2014ascl.soft08004N} to have a minimum of 10 counts per channel. We used the $\rm \chi^2$ test statistic and performed spectral analysis in the 0.3--8 keV energy band. To compute 90$\%$ confidence intervals for the best-fit parameters, we used the {\tt error} tool in XSPEC. We used the Tuebingen-Boulder ISM absorption model ({\tt tbabs} in XSPEC,  the solar elemental abundances from \citealt{2000ApJ...542..914W}) to account for interstellar absorption. The unabsorbed fluxes were computed by using the {\tt cflux} convolution model in XSPEC.

We initially approximated the X-ray spectrum with power-law ({\tt powerlaw}) and optically thin thermal emission models ({\tt mekal} with fixed metal abundance at solar value). Both power-law ($\rm \Gamma = 1.27$) and mekal (${\rm k}T\ga 15$ keV) models give an acceptable fit to our spectrum, resulting in reduced chi-square ($\rm \chi^2_{red}$) values of 0.99 and 0.98, respectively (see Table \ref{tab:xfit}). We used an isobaric cooling flow model ({\tt mkcflow}; \citealt{1988ASIC..229...53M}) to fit the X-ray spectrum of Gaia22ayj. The {\tt mkcflow}\footnote{To properly use the {\tt mkcflow} model, the redshift parameter can not be set at zero value (see for more details \citealt{2017mukai}). We fixed the redshift parameter at $\rm 5.84 \times 10^{-7}$ (equal to the 2,500 pc distance) using the cosmological Hubble constant of $\rm 70\ km\ s^{-1}\ Mpc^{-1}$.} provides a good approximation of the X-ray spectrum of non-magnetic CVs \citep[e.g.,][]{2017mukai}. The single fit with the {\tt mkcflow} results in an unacceptable fit ($\rm \chi^2_{red}=1.12$) and gives only the lower limit for the temperature (${\rm k}T\ga 64$ keV). Along with the resulting photon index of the power-law model ($\rm \Gamma = 1.27$), this indicates that the X-ray spectrum of Gaia22ayj is hard. X-ray spectra of some magnetic CVs cannot be fitted well with the single {\tt mkcflow} model, showing a steep photon index ($\rm \Gamma \sim 1$) for a power-law model \citep[e.g.,][]{2017mukai,2021AstL...47..587G}. The X-ray spectrum of magnetic CVs might be affected by local absorbers, so we added the partial covering absorption component ({\tt pcfabs}) to the {\tt mkcflow} model \citep[e.g.][]{2017mukai}. The final model $\tt tbabs\times pcfabs\times mkcflow$ approximates well the observed spectrum of Gaia22ayj and gives an acceptable fit ($\rm \chi^2_{red}=1.01$; see Figure \ref{fig:Xray_spectrum}). 

The X-ray spectrum of some polars shows a soft component along with the thermal plasma emission model \citep[e.g.][]{2017mukai}. We approximated the X-ray spectrum by including the black-body component ({\tt bbody}) in the final model, having $\tt tbabs*pcfabs*(mkcflow+bbody)$. However, the fit gives no meaningful result for the black-body temperature.

Table \ref{tab:xfit} shows the best-fit spectral parameters for different models. The hydrogen column density, $N_\textrm{H}$, from our fit is consistent with the Galactic value for Gaia22ayj from the Bayestar dust map ($\rm \sim 2\times10^{20}\ cm^{-2}$;E(g--r) = 0.03;\citealt{bayestar19}). To compute the bolometric flux in the 0.001--100 keV energy band, we used the {\tt cflux} model for the {\tt mkcflow} component. We converted the fluxes into luminosities by assuming the 2.5 kpc \textit{Gaia}-estimated distance to the object. Gaia22ayj shows an observed X-ray luminosity of $\rm \approx 2.5\times 10^{32}$ erg~s$^{-1}$ and a bolometric X-ray luminosity of $\rm \approx 4\times 10^{32}$ erg~s$^{-1}$. We extracted the X-ray spectrum for 11 Swift/XRT observations to find possible X-ray variability. The current analysis shows no change of spectral parameters between six months of the observation with Swift/XRT (from June 21, 2005 to December 6, 2005). 

\subsubsection{X-ray Timing Analysis}
Finally, we performed a timing analysis on the combined Swift data. This required downloading the raw data and performing a barycentric correction, converting all photon arrival times to BJD$_\textrm{TDB}$. We used the Gregory-Loredo algorithm \citep[GL;][]{1992gl}, which is a phase-folding Bayesian algorithm that performs well in the regime of low to moderate counts, which is common in X-ray data like that which we present here. For a full explanation of the implementation and application of this algorithm to other sources, we refer the reader to \cite{2023bao1, 2024bao2}.  In Figure \ref{fig:Xray_spectrum}, we show the folded light curve at the best X-ray period obtained with the GL algorithm: 9.635(2) min. We also show the X-ray light curve folded on the optical period (9.36 min), which shows that modulation is still detected on that period, though not as clear and significant as on the 9.64-min period. However, we strongly emphasize that the period we obtain has significant overlap with harmonics of the ``Good Time Interval'' (GTI) of the \textit{Swift} observation, meaning that this analysis is subject to scrutiny and further X-ray observations and timing analyses should be done to confirm this period.

\subsection{Radio Observation}
\begin{figure}
    \centering
    \includegraphics[width=0.5\textwidth]{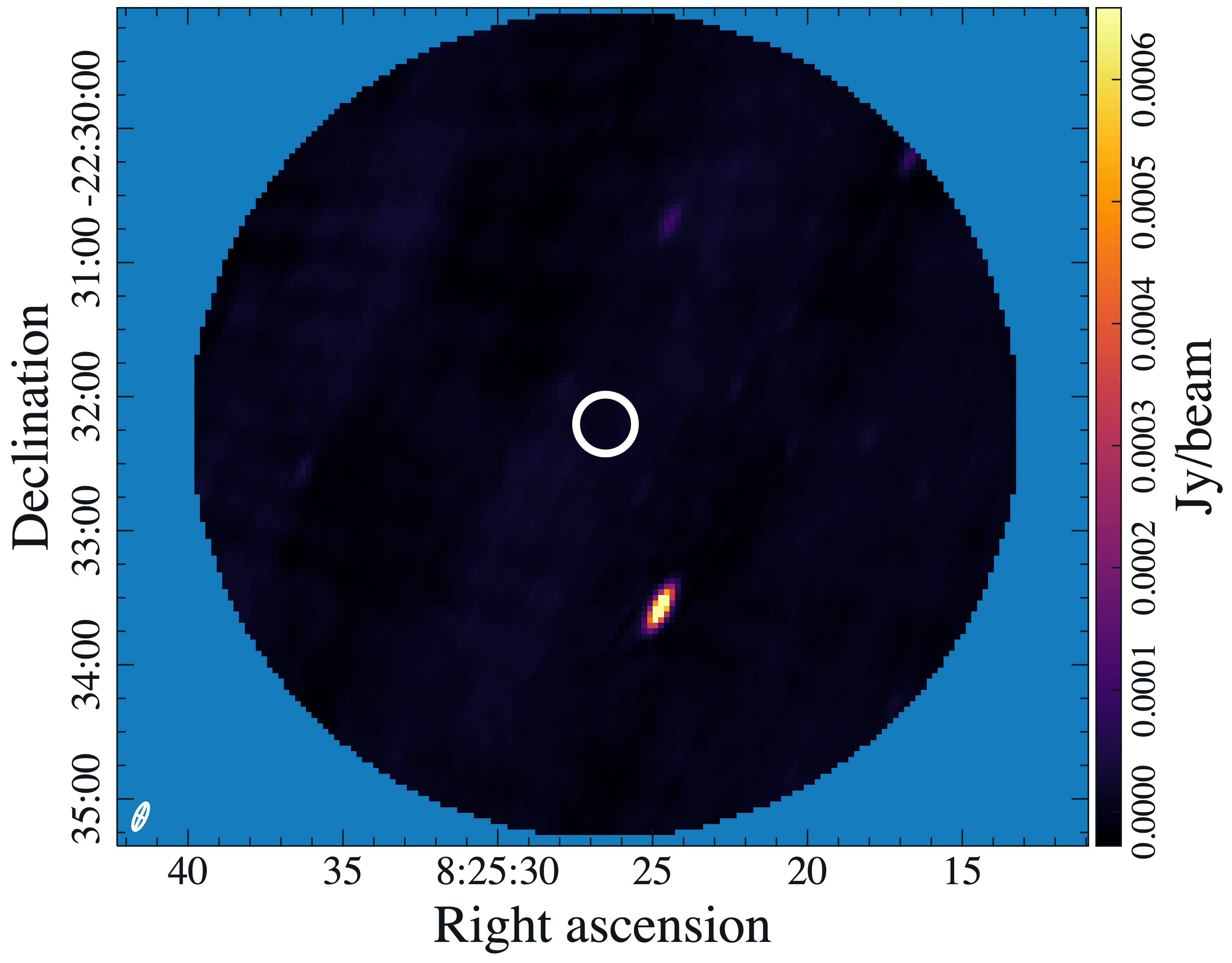}\\
    \includegraphics[width=0.45\textwidth]{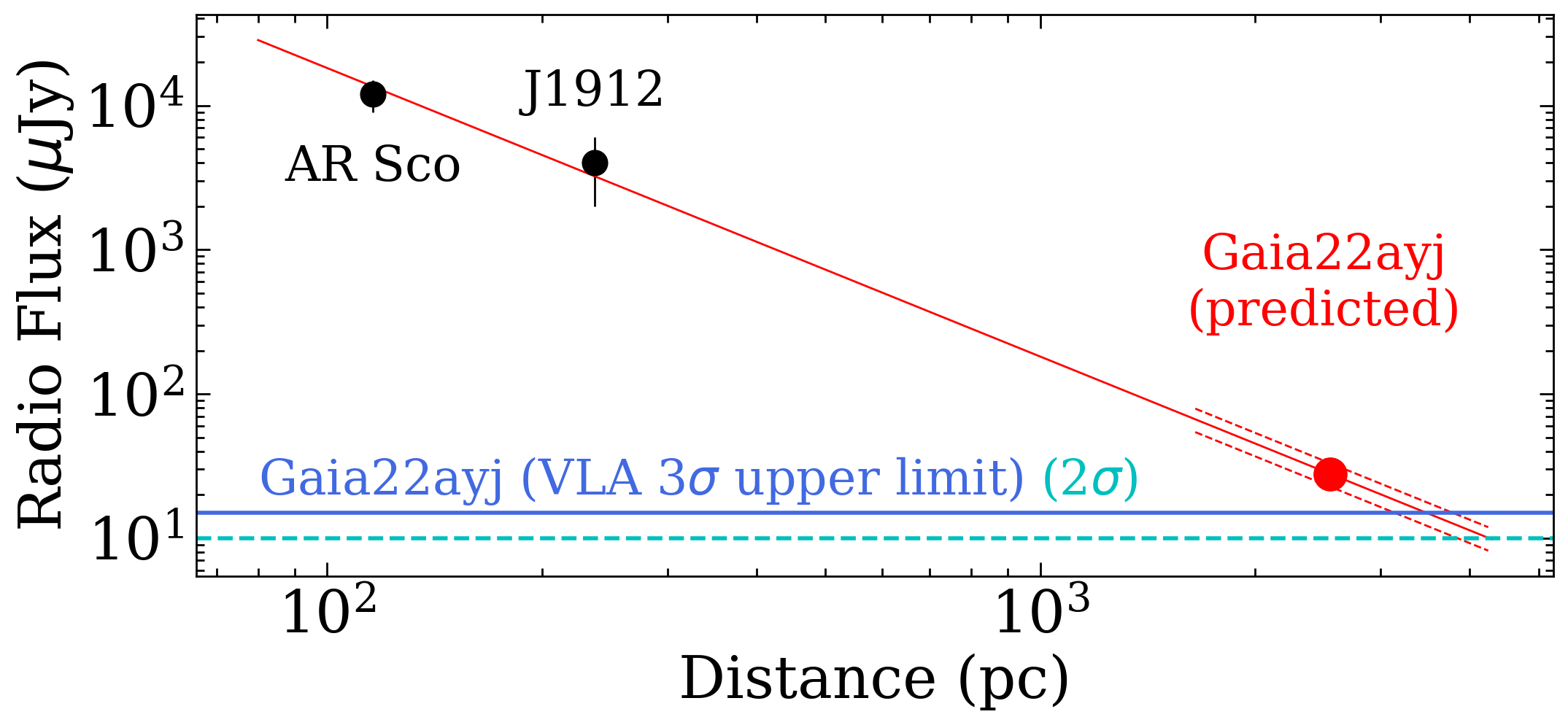}
    \caption{\textit{Top: }VLA non-detection (3$\sigma$ upper limit of 15 $\mu$Jy) of Gaia22ayj (a 6$\arcsec$ radius white circle is shown around the optical position). An unassociated radio source is located approximately 1.45$\arcmin$ to the south west of the field center, with the synthesized beam shown on the lower left. \textit{Bottom: } If the same radio emission mechanism were present as in the known WD pulsars, AR Sco and J1912, we should have seen a radio flux of $F_\nu = 30^{+50}_{-20} \;\mu\textrm{Jy}$ from Gaia22ayj. VLA observations rule this out.}
    \label{fig:radio}
\end{figure}

We observed Gaia22ayj with the Very Large Array (VLA) for 3 h on two separate occasions: 13 and 15 Jan, 2024 (UT), both in X-band (8--12 GHz) with the VLA in the ``D" array configuration. Weather conditions were good (at most 20\% cloud cover) on both occasions. We used the flux calibrator 3C138 and the
gain calibrator J0826--2230. Data were calibrated using the
Common Astronomy Software Applications software \citep[CASA;][]{casa} and deconvolved using the \texttt{clean} algorithm with the \texttt{tclean} command. We
measured the flux density using the Cube Analysis and Rendering Tool for Astronomy \citep[CARTA;][]{carta} and found no detection in either the dirty or deconvolved image. To deconvolve the image, we used the \texttt{clean} algorithm \citep{clean} within CASA. The rms at the source position was 5.0 $\mu$Jy, which corresponds to a 3$\sigma$ upper limit of 15 $\mu$Jy. We show the cleaned image in Figure \ref{fig:radio}, which shows no significant source in the vicinity of Gaia22ayj, but reveals an unassociated radio source (likely a radio galaxy) 1.45' to the south west.

In Figure \ref{fig:radio}, we show that we expected Gaia22ayj to have a radio flux of $F_\nu = 30^{+50}_{-20} \;\mu\textrm{Jy}$. This was calculated based on a least-squares fit to the radio fluxes of AR Sco \citep{2016marsh} and J1912 \citep{2023pelisoli_j1912}, assuming that Gaia22ayj was powered by the same physical mechanism. The uncertainties on our estimates of the radio flux of Gaia22ayj are large due to the distance uncertainty from \textit{Gaia}. Figure \ref{fig:radio} shows that the radio non-detection of Gaia22ayj is inconsistent with it having a similar radio luminosity as the known WD pulsars at the 3$\sigma$ (2$\sigma$) level, assuming a distance of 2.5 kpc (4 kpc).

\subsection{Near-Infrared Spectroscopy}
We observed Gaia22ayj on 30 April 2023 with the  Near-Infrared Echellette Spectrometer (NIRES) on the Keck II telescope on Mauna Kea. We obtained an ABBA dither sequence of four 180-s exposures. Spectra were reduced using standard techniques with the NSX pipeline\footnote{\url{https://sites.astro.caltech.edu/~tb/nsx/}}. We find a significant detection of the He\,{\sc i} emission line at 1.083 $\mu$m, but no other discernible lines above the sky background.  


\subsection{Gamma-Ray False Association}
Gaia22ayj is located 7.91$\arcmin$ away from the well-studied BL Lacertae source PKS 0823--223 \citep{1982allen}, also known as 4FGL J0825.9--2230 in the \textit{Fermi} 4FGL catalog \citep{4fgl}. Gaia22ayj is within the  $27.5\arcmin  \times 26.6\arcmin$ \textit{Fermi} error ellipse (68\% confidence intervals). However, Gaia22ayj can be confidently discarded as being associated with the gamma-ray source since the 4FGL catalog reports a $>$99.9\% likelihood of PKS 0823--223 being associated with the gamma-ray source. A search for gamma-ray pulsations on the 9.36-min period could still be conducted, but is beyond the scope of this work.

\section{Results and Interpretation}
\label{sec:analysis}
\subsection{An Accreting Binary System}
\cite{2022kato} first suggested Gaia22ayj to be an eclipsing double WD system, but the optical spectrum in Figure \ref{fig:spec} shows clear emission lines. No absorption lines are seen anywhere in the spectrum at any orbital phase, which provides clear evidence against a \textit{detached} double WD system. Furthermore, Figure \ref{fig:spec} shows that the He\,{\sc ii} 4686 and H$\alpha$ lines are double-peaked, and have a characteristic broadening of $v\approx 1200~\textrm{km s}^{-1}$. This discards the possibility of a chromospheric origin of the emission lines (which would be broadened at the 10-20$~\textrm{km s}^{-1}$ level), meaning that accretion must be taking place in a semi-detached binary system (i.e. one star fills its Roche lobe). 

\subsection{The 9.36-min Period: Not an Ultracompact System}
In the absence of additional information, the observed 9.36-min period could either be attributed to the binary orbit or to the spin of the accreting star. However, there is also the possibility that this is the beat period between the spin and the orbit (i.e. $1/P_\textrm{beat} = 1/P_\textrm{spin} \pm 1/P_\textrm{orbit}$). The spectrum in Figure \ref{fig:spec} shows that this cannot be an orbital period because the spectrum is dominated by hydrogen emission lines and a clear Balmer jump in emission (inverse Balmer jump) at 3645 \AA. 

CVs have an observed orbital period minimum of approximately 78--82 minutes \citep{2009gaensicke}, which has been explained by calculating the minimum orbital period that a hydrogen-rich donor star can remain in thermal equilibrium and continue hydrogen burning while filling its Roche lobe \citep[e.g.][]{1983paczynski}. However, this orbital period minimum can reach $\approx 51$ minutes for the smallest, lowest metallicity stars \citep{1997stehle}. Even more generally, a degenerate companion object (i.e. a gas giant planet or a brown dwarf) can reach orbital periods as low as 37 minutes \citep{2021rappaport}. 

Gaia22ayj, with a photometric period of 9.36 min, is well below this limit and shows no evidence of helium-rich accretion that would suggest a high-density donor\footnote{It is worth nothing that though hydrogen has been seen in ultraviolet spectra of ultracompact (AM CVn) WD binaries \citep[such as the 5.4-min orbital period HM Cnc;][]{2023munday}, the amount of hydrogen is estimated to be very low, meaning optical spectra are still dominated by helium lines.}, so the observed period \textit{must} be related to the spin period of the WD. The overall modulation of the spectrum on this period and the lack of emission-line RV shifts (Figure \ref{fig:spec}) also support this claim. 

\subsection{Spectral Energy Distribution and Upper Limits on the Donor Star}

\begin{figure}
    \centering
    \includegraphics[width=0.5\textwidth]{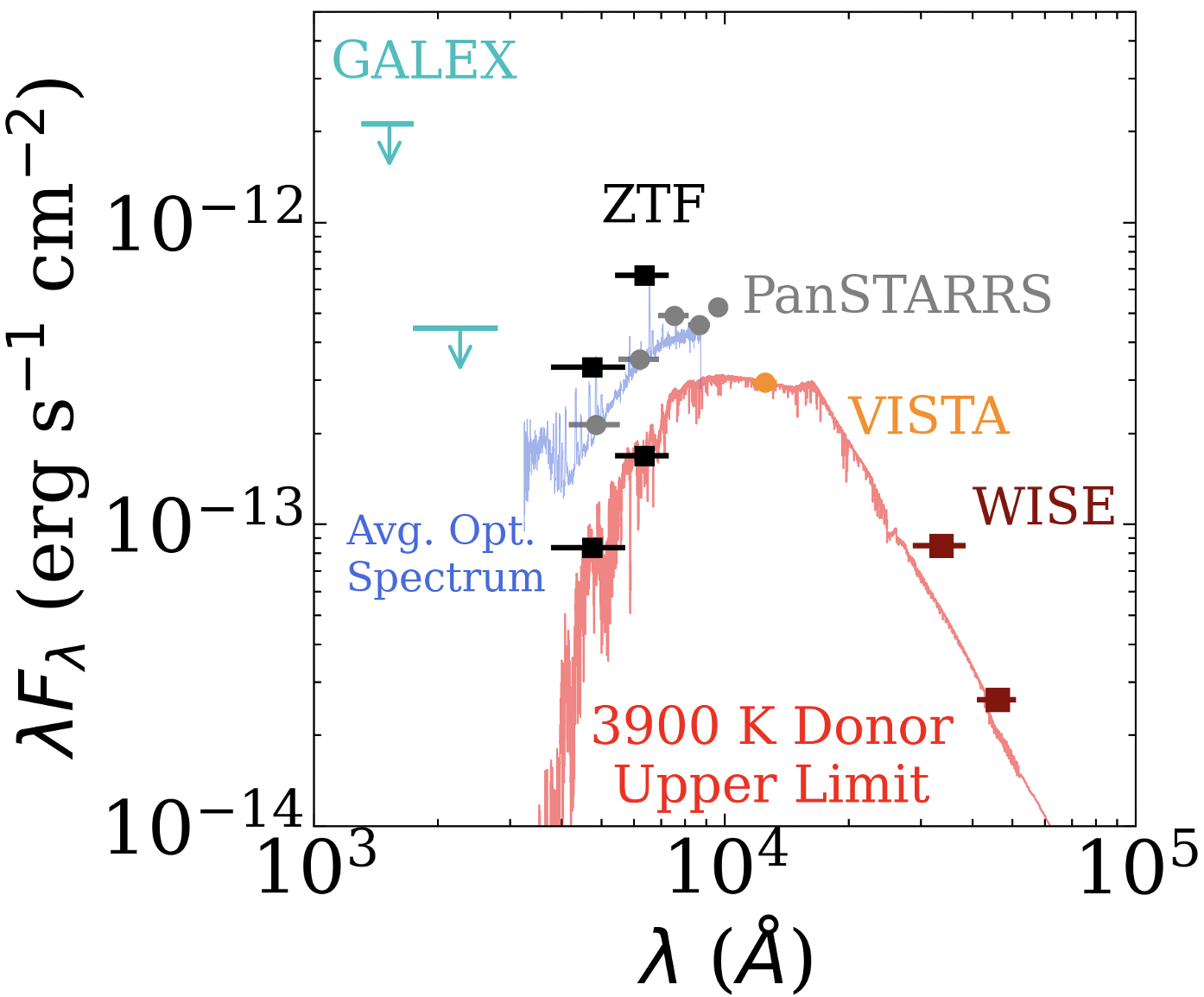}
    \caption{A 3900-K donor, with $R_\textrm{donor} = 0.62~R_\odot$ best fits the optical light curve minimum of Gaia22ayj. This is consistent with near-IR and mid-IR photometry from VISTA and WISE, respectively, and allows us to place upper limits on the orbital period of Gaia22ayj (Figure \ref{fig:cartoon}).}
    \label{fig:sed}
\end{figure}

In Figure \ref{fig:sed}, we show the interstellar extinction-corrected spectral energy distribution (SED) of Gaia22ayj. We use the $E(g-r)$ value of 0.03 from \cite{bayestar19}, which, given the Galactic coordinates of Gaia22ayj ($\ell$: 243.93762$^\circ$, $b:$ 8.80238$^\circ$) where interstellar extinction is low, should be a reliable estimate.  UV (GALEX GR6/7; upper limits), optical (PanSTARRS PS1), near-IR (VISTA), and mid-IR (WISE) points are shown, along with the phase-averaged Keck I/LRIS spectrum. ZTF photometry is plotted at the light curve maximum and minimum. The minimum of the ZTF photometry allows us to place an upper limit on the temperature of the donor star. We take a grid of stellar atmospheres at a given temperature (at solar metallicity) from the BT-Settl library \citep{btsettl} and corresponding stellar radii from the models of \cite{2011knigge}, and compare to the ZTF $g$ and $r$ points at light curve minimum. Assuming a donor behaving according to the \cite{2011knigge} tracks (i.e. not evolved), we find that a donor with $T_\textrm{eff} = 3900$ K, $R_\textrm{donor} = 0.62~R_\odot$ best fits the observed limits, given the median \textit{Gaia} distance of $d\approx 2500$ pc. In Section \ref{sec:link}, we elaborate on how this finding could inform the potential orbital period of the system and its evolutionary stage.

\subsection{Polarization Confirms a Magnetic White Dwarf}
Linear polarization in Gaia22ayj confirms the magnetic nature of the object. Most interestingly, the only system to date that has shown such high levels (40\%) of linear polarization is AR Sco, the prototypical WD pulsar, which also showed up to 40\% linear polarization \citep{2017buckley}. In AR Sco, the linear polarization is in phase with the total optical modulation, whereas in Gaia22ayj (Figure \ref{fig:hipercam}), the two are clearly in antiphase. 

Polars are typically polarized to at most a few percent in linear polarization, and can only reach high levels of \textit{circular} polarization \citep[e.g.][]{1992Schaich}. \textit{No} IP is known to exhibit linear polarization at the level observed in Gaia22ayj \citep[e.g.][]{2020ferrario}, suggesting that this system has a high magnetic field and could have at most a small truncated disk. Two distinct peaks in linear polarization percentage and a swing in polarization angle across spin phase, however, may suggest the presence of two accretion poles, as famously seen in the prototypical polar CV, AM Her, though circular polarization will tell for certain \citep{1991amher_pol}.

This may also be supported by the nearly double-peaked nature of the He\,{\sc ii} 4686 and H$\alpha$ emission lines, that remain constant throughout the spin phase (Figure \ref{fig:spec}) and even during $\approx 2$-h long integrations (Figure \ref{fig:total_spec}). This suggests that either 1) a disk is present or 2) that we are always viewing down one accretion pole in a two-pole system, with the second pole behind the WD also contributing to the observed emission. The high velocity of each peak ($v\approx1200~ \textrm{km s}^{-1}$ broadening) suggests that in the former scenario, those lines trace matter in a disk in Keplerian rotation at $r \approx 0.1\,R_\odot$ around a $0.8\,M_\odot$ WD, or in the latter scenario, material in free-fall down the accretion stream traveling at that speed. The absence of RV shifts in the emission lines slightly favors the two-pole scenario, as the disk around the WD in the first scenario would display some motion relative to the center of mass. In the two-pole accretion stream scenario, if a pole is viewed head-on, the emission region would exhibit minimal motion as the WD rotates.

\subsection{Optical Spectroscopy Suggests a High Magnetic Field Strength}
The high-amplitude optical modulation seen in the spectrum (see Figure \ref{fig:spec}, upper left panel) is reminiscent of cyclotron ``humps" in polar CVs. Setting the magnetic force equal to the centripetal force leads to the derivation of cyclotron harmonics:
\begin{gather}
    \lambda = \frac{10710\,\textrm{\AA}}{n}\left(\frac{100\,\textrm{MG}}{B}\right)\sin\theta~, 
    \label{eq:cyc}
\end{gather}
where $n$ is the cyclotron harmonic number, $B$ the WD magnetic field strength, and $\theta$ the angle of the magnetic pole with respect to our line of sight. 

The following two observations are critical for determining the magnetic field strength of the WD: 1) individual cyclotron humps are not observed, and 2) there is high-amplitude optical modulation, with two distinct ``humps" spanning nearly the entire optical spectrum peaking at \textbf{spin} phases 0.25 and 0.75 (Figure \ref{fig:spec}). The former provides a clue that the magnetic field should be weak enough such that only high harmonics should be present at optical wavelengths. Cyclotron modeling and observations reveal that $n\gtrsim7$ harmonics are difficult to discern as single features \citep[e.g.][]{2008campbell}. The latter suggests that the viewing angle must be high enough for high amplitude modulation to be seen \citep[e.g.][]{2008campbell}. Therefore, we set $\theta=60^\circ$ in Equation \ref{eq:cyc} as a lower limit for the viewing angle. Because the $n=7$ feature is not seen at $\lambda < 8800\,{\rm \AA}$, we infer an upper limit on the WD magnetic field strength of $B \lesssim15$ MG (Equation \ref{eq:cyc}).


In order to place a lower limit on $B$, we assume that $n>20$ harmonics must be present at optical wavelengths, which implies $B\gtrsim 5$ MG. The lowest measured magnetic field of a polar CV showing Zeeman splitting has been 7 MG in V2301 Oph \citep{1995ferrario}, though its optical variability (out of eclipse) does not exceed 0.5 mag in ZTF. This shows that our 5 MG lower limit is quite conservative, and that Gaia22ayj likely has a stronger magnetic field. 

\subsection{Is the Accretion Stable?}
The magnetospheric radius is the point where the ram pressure of infalling material equals the magnetic pressure due to the WD. At this point, accreted matter is forced to follow the magnetic field lines of the WD, and is channeled onto its surface. The magnetospheric radius is:
\begin{gather}
    r_A = \left(\frac{B^4R^{12}}{2 G M \dot{M}^2}\right)^{1/7} \\
    = 0.62 R_\odot \left(\frac{B}{10 \textrm{ MG}}\right)^{4/7}\left(\frac{R}{0.0105\,R_\odot}\right)^{12/7}\times \notag\\
    \left(\frac{M}{0.8\,M_\odot}\right)^{-1/7}\left(\frac{\dot{M}}{10^{-9}\,M_\odot \textrm{ yr}^{-1}}\right)^{-2/7} \notag~,
\end{gather}
where $M$ is the WD mass, $\dot{M}$ is the accretion rate, $B$ is the WD field strength, and $R$ is the WD radius. As matter makes its way to the magnetospheric radius, two possible scenarios can occur. If the centrifugal force on a piece of matter exceeds the gravitational force from the WD, then it will be flung out, removing angular momentum from the system (i.e. a ``propeller" phase). Otherwise, it will be accreted onto the WD, transferring angular momentum to the WD. This leads to a condition of ``spin equilibrium" \cite[e.g.][]{1994patterson}, and involves finding the orbital period of material at the magnetospheric radius:
\begin{gather}
\label{eq:spin}
    P_\textrm{eq}^2 =\frac{4\pi^2r_A^3}{GM} = \frac{4\pi^2}{GM}\left(\frac{B^4R^{12}}{2 G M\dot{M}^2}\right)^{3/7}\\
    \Longrightarrow P_\textrm{eq} = 11.7 \textrm{ min }\left(\frac{B}{10 \textrm{ MG}}\right)^{6/7}\left(\frac{R}{0.0105\,R_\odot}\right)^{18/7}\times \notag\\
    \left(\frac{M}{0.8\,M_\odot}\right)^{-17/14}\left(\frac{\dot{M}}{10^{-9}\,M_\odot \textrm{ yr}^{-1}}\right)^{-3/7} \notag~.
\end{gather}

\begin{figure}
    \centering
    \includegraphics[width=0.45\textwidth]{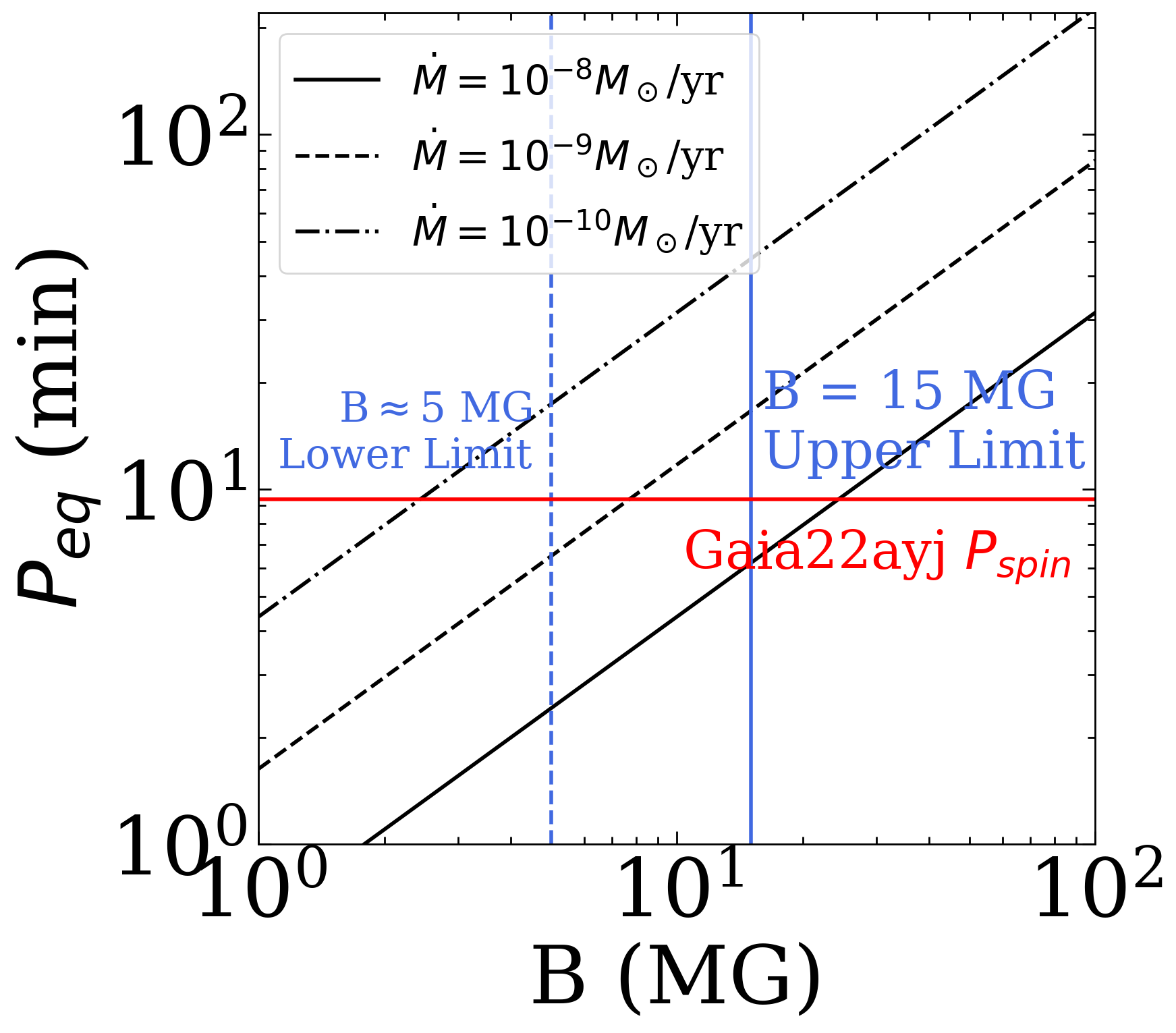}
    \caption{Equilibrium spin period as a function of magnetic field strength, for different accretion rates (Equation \ref{eq:spin}). Limits on magnetic field strength from Equation \ref{eq:cyc} are shown as vertical blue lines, and the 9.36-min spin period as a horizontal red line. Given the magnetic field constraints, $\dot{M} \gtrsim 5\times10^{-10}\,M_\odot \textrm{ yr}^{-1}$ is required for $P_\textrm{eq} < P_\textrm{spin}$, the condition required for stable accretion.}
    \label{fig:equilibrium}
\end{figure}

We plot the result of this analysis in Figure \ref{fig:equilibrium}. In order to sustain accretion in such a fast-spinning WD, either the magnetic field must be weak or the accretion rate must be high. In IPs, the 1--10 MG magnetic field allows for stability across a wide range of accretion rates: low-luminosity IPs \citep[LLIPs;][]{2014pretorius} are below the CV orbital period gap, and are thought to have $\dot{M} \lesssim10^{-10}\,M_\odot\textrm{ yr}^{-1}$, while typical IPs are above the gap and accrete at $\dot{M} \gtrsim 10^{-9}\,M_\odot\textrm{ yr}^{-1}$ \citep[e.g.][]{2019suleimanov}. 

We proceed with the assumption that Gaia22ayj is indeed accreting (i.e. not in a propeller phase). This is justified by the following: 1) the X-ray luminosity exceeds that of AE Aqr, the prototypical propeller, by at least an order of magnitude \citep{1980patterson}; 2) there is no clear optical flaring in the light curve, unlike AE Aqr and LAMOST J0240, the two known WD propellers \citep{2020thorstensen}; and 3) the spin period of Gaia22ayj is twenty times longer than that of the two known propellers \citep[33 and 24 seconds;][]{1979patterson, 2022pelisoli}.

In the case of Gaia22ayj, we know that $B\gtrsim5$ MG in order for such strong cyclotron emission and polarization to be observed. In order for stable mass transfer to be taking place, Figure \ref{fig:equilibrium} suggests that $\dot{M} \gtrsim 5\times10^{-10}\,M_\odot \textrm{ yr}^{-1}$. Since this is a conservative lower limit, it is likely that $\dot{M} \gtrsim 10^{-9}\,M_\odot \textrm{ yr}^{-1}$, placing Gaia22ayj above the CV orbital period gap, as explored further in Section \ref{sec:link}.

\subsection{A White Dwarf Rapidly Spinning Down}
\begin{figure}
    \centering
    \includegraphics[width=0.48\textwidth]{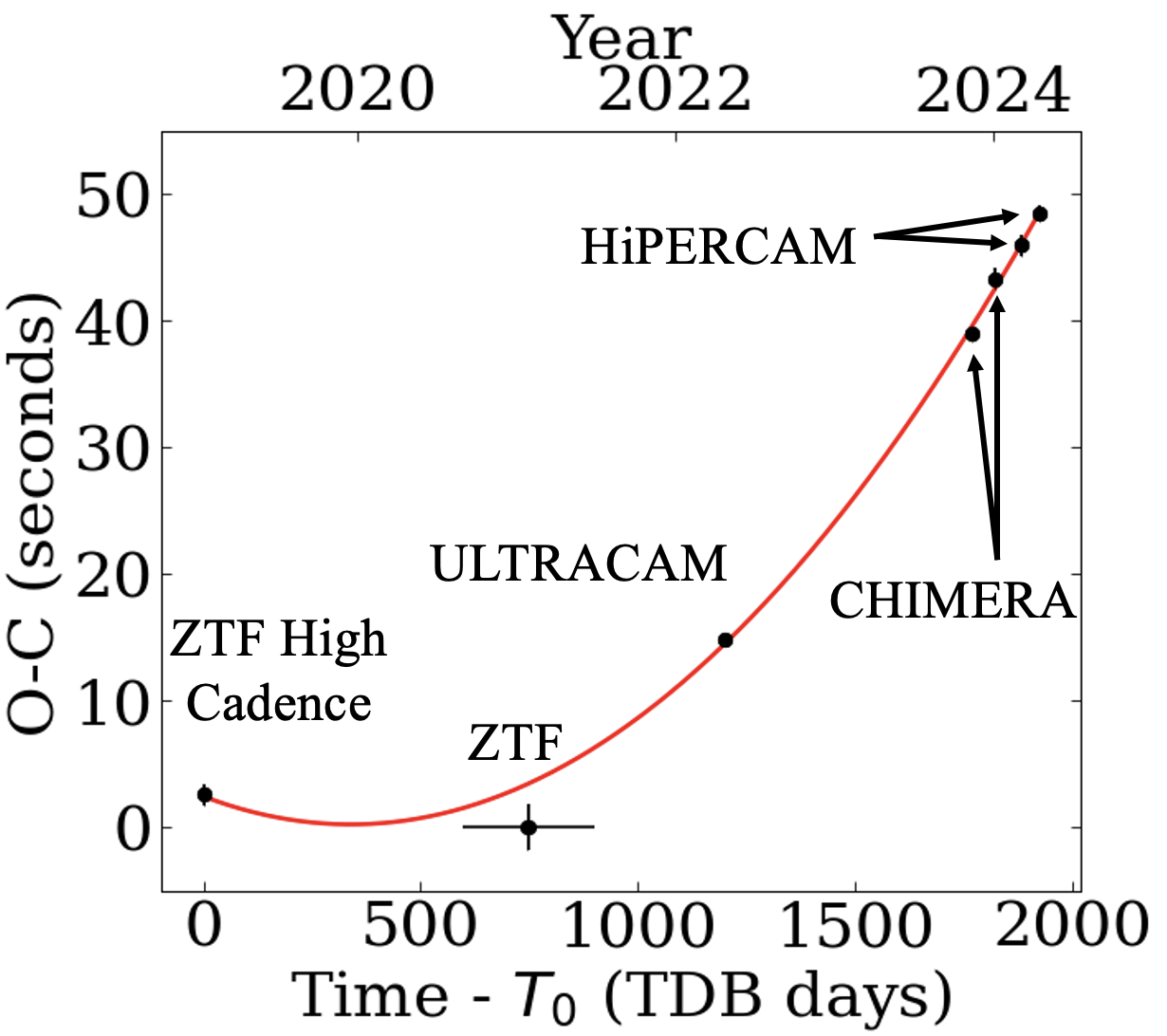}\\
    \includegraphics[width=0.48\textwidth]{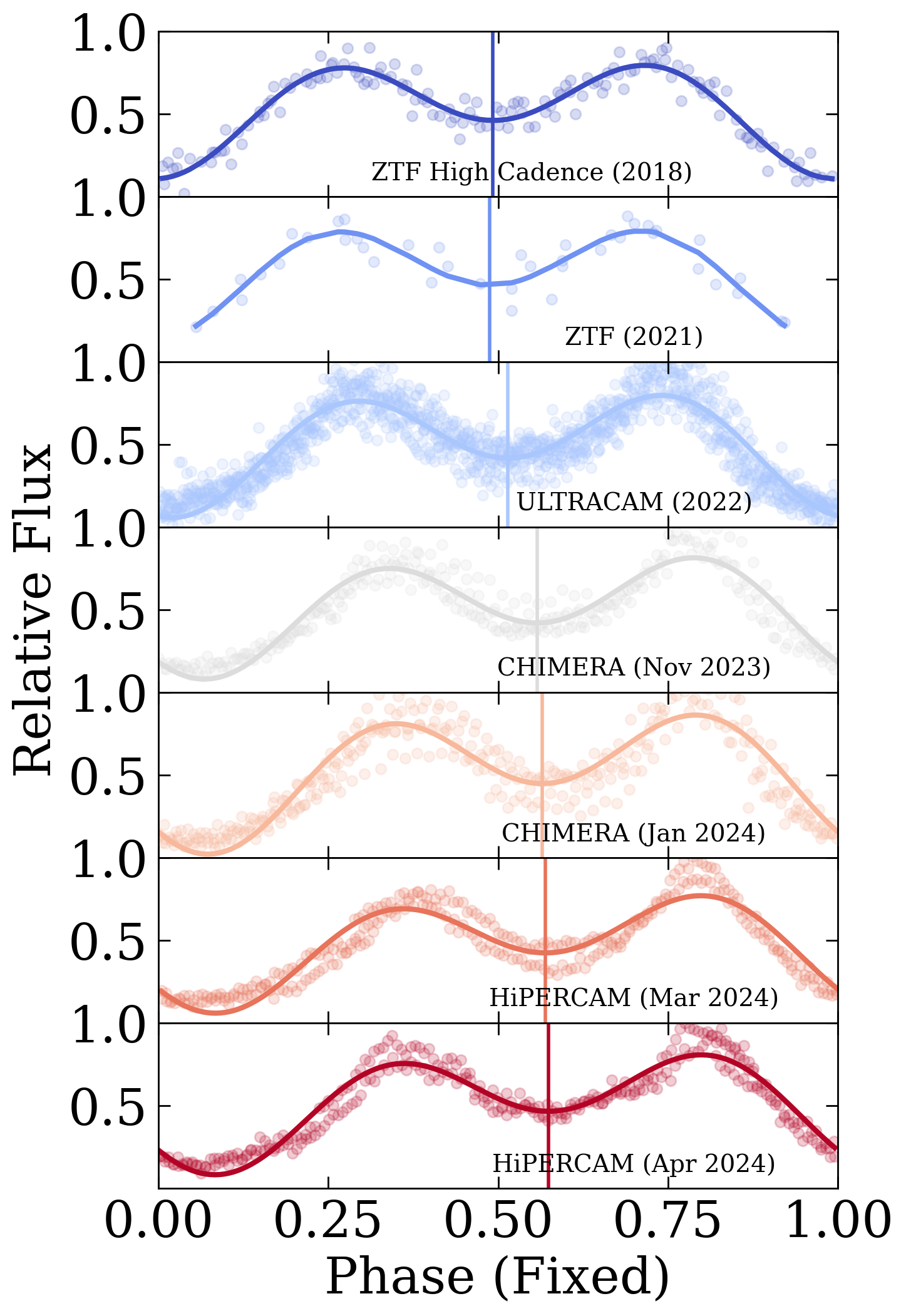}
    \caption{The ``observed" minus ``expected" (O--C) diagram of Gaia22ayj (top) shows that the expected time of the light curve minimum has drifted over six years. A multiyear, high-speed optical photometry campaign demonstrated that Gaia22ayj is spinning down at $\dot{P} = (2.89\pm 0.12)\times 10^{-12}\; \textrm{s s}^{-1}$, about four times higher than AR Sco, though with a similar characteristic timescale $P/\dot{P}\sim 5\times10^6$ yr. }
    \label{fig:pdot}
\end{figure}

We folded the entire ZTF light curve of Gaia22ayj, with data between 2018 and 2024, and obtained the best period using a Lomb-Scargle analysis \citep{1976lomb, 1982scargle}. Based on that, we then computed an ``observed minus calculated" ($O-C$) value for each high-speed light curve obtained between 2018 and 2024. We fit a two-component sinusoid to the ZTF $r$-band light curve and fit the same model to all other high-speed light curves. We computed the deviation of the light curve minimum (phase 1.0 in Figure \ref{fig:main}) from the original ZTF template, and created the $O-C$ diagram in Figure \ref{fig:pdot}. We also show all high-speed light curves along with their best-fit two-component sinusoid, and show how one light curve minimum (phase 0.5) drifts from the original ZTF fixed phase in Figure \ref{fig:pdot}.

We then measure $\dot{P}$ by obtaining the best-fit (least squares) quadratic function (red line) to the data in Figure \ref{fig:pdot}. $\dot{f}$ is given (e.g., Equation 1 in \citealt{2019burdge}) by:
\begin{gather}
    \Delta t_\textrm{O--C} = \frac{1}{f_0}\left(\frac{1}{2}\dot{f_0}(t - t_0)^2 + \frac{1}{6}\ddot{f_0}(t - t_0)^3 + \cdots\right)~,
\end{gather}
where $t$ is an epoch of observation, $t_0$ the initial epoch, $\Delta t_\textrm{O--C}$ the measured deviation in the light curve, $f_0$ is the frequency at $t_0$, $\dot{f_0}$ its first derivative, and $\ddot{f_0}$ its second derivative. Since $\ddot{f}$ is consistent with zero, we only fit a parabola to the observed data. We calculate $\dot{f_0}$ directly, assume that it remains constant throughout the entire span and estimate errors based on the covariance matrix of the best-fit parameters: $\dot{f} = (2.57 \pm 0.11) \times 10^{-15}\,\textrm{Hz s}^{-1}$. Propagating errors to measure $\dot{P}$, we obtain $\dot{P} = (2.89\pm 0.12)\times 10^{-12}\; \textrm{s s}^{-1}$. 

\begin{figure}
    \centering
    \includegraphics[width=0.48\textwidth]{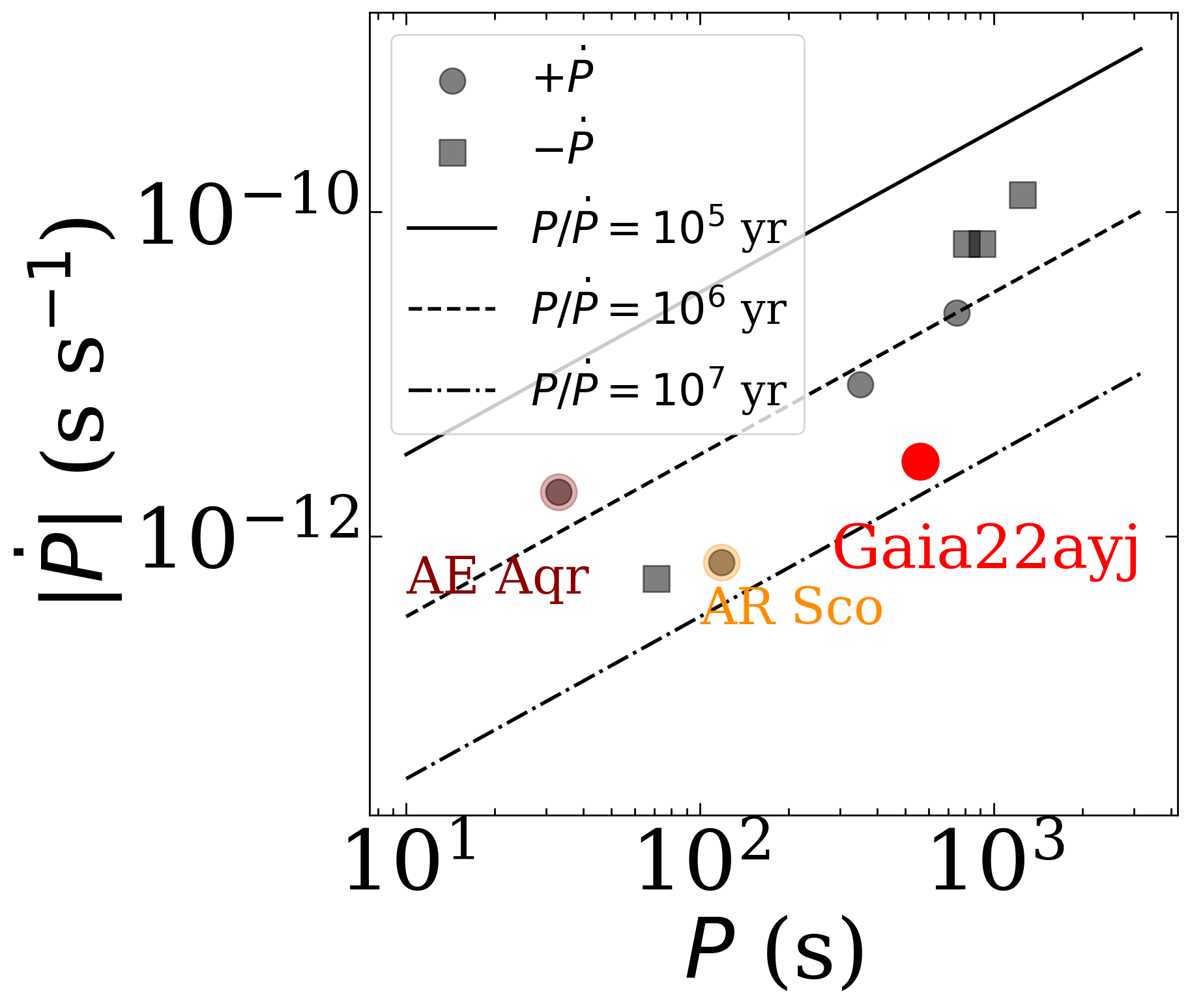}
    \caption{IPs that show consistent spin up (squares) or spin down (circles), including the propeller AE Aqr, are shown on the $P-\dot{P}$ diagram. AR Sco and Gaia22ayj are related in having the longest characteristic spin-down times of known systems.}
    \label{fig:pdot_space}
\end{figure}

In Figure \ref{fig:pdot_space}, we compare the measured characteristic spin-down timescale of Gaia22ayj ($P/\dot{P} \approx 6.1^{+0.3}_{-0.2}\times 10^6$ yr) to that of other IPs, the propeller system AE Aqr, and the WD pulsar AR Sco. The IPs used as reference are systems that have a consistent spin up/down as determined by \cite{2020patterson}: DQ Her, FO Aqr, V1223 Sgr, BG CMi, and GK Per. AE Aqr and AR Sco have separately measured period derivatives, determined by \cite{1994dejager} and \cite{2022pelisoli_arsco}, respectively. Figure \ref{fig:pdot_space} shows that no IPs have as high a characteristic spin-down timescale ($\tau = P/\dot{P}$) as AR Sco ($5.6\times 10^6$ yr) and Gaia22ayj ($6.1\times 10^6$ yr). Curiously, the spin-down timescale of AE Aqr is comparable to that of other IPs.

Finally, in Table \ref{tab:all_info}, we summarize all observed and inferred parameters of Gaia22ayj, the latter of which are obtained from arguments and analysis presented above.

\begin{table}[]
    \centering
    \begin{tabular}{l|c|c}
         Quantity & Value & Source \\
         \hline
         RA (hms) & 08:25:26.52 & \textit{Gaia} DR3\\
         DEC (dms) & -22:32:12.34 & \textit{Gaia} DR3\\
         Distance (kpc) & 2.5$^{+1.5}_{-1.0}$ & \textit{Gaia} DR3\\
         $G$ (mag) & 19.2 & \textit{Gaia} DR3\\
         $BP - RP$ (mag) & 1.11 & \textit{Gaia} DR3\\
         Gaia DR3 ID & 5697000580270393088 & \textit{Gaia} DR3\\
         $F_X$ ($\textrm{erg s}^{-1} \textrm{cm}^{-2}$) & $(3.9 \pm 0.9) \times 10^{-13}$& Swift/XRT\footnote{0.3--8 keV}\\
         $F_r$ ($\mu$Jy) & $<15.8$ ($3\sigma$ upper lim.) & VLA\footnote{X-band (8--12 GHz)}\\
         $P_\textrm{opt}$ (min) & 9.3587(1) & ZTF\\
        $P_\textrm{X-ray}$ (min) & 9.635(2) & Swift/XRT\footnote{Subject to scrutiny due to possible beat with Swift GTI}\\
         $\dot{P}_\textrm{opt} (\textrm{s s}^{-1})$&$(2.89\pm 0.12)\times 10^{-12}$ & ZTF\\
         $P/\dot{P}$ (yr) &$6.1^{+0.3}_{-0.2}\times 10^6$ & ZTF\\
\hline
        
    \end{tabular}
    \caption{Summary of observed properties of Gaia22ayj. }
    \label{tab:all_info}
\end{table}

\section{Discussion}
\label{sec:discussion}
\begin{figure}
    \centering
    \includegraphics[width=0.45\textwidth]{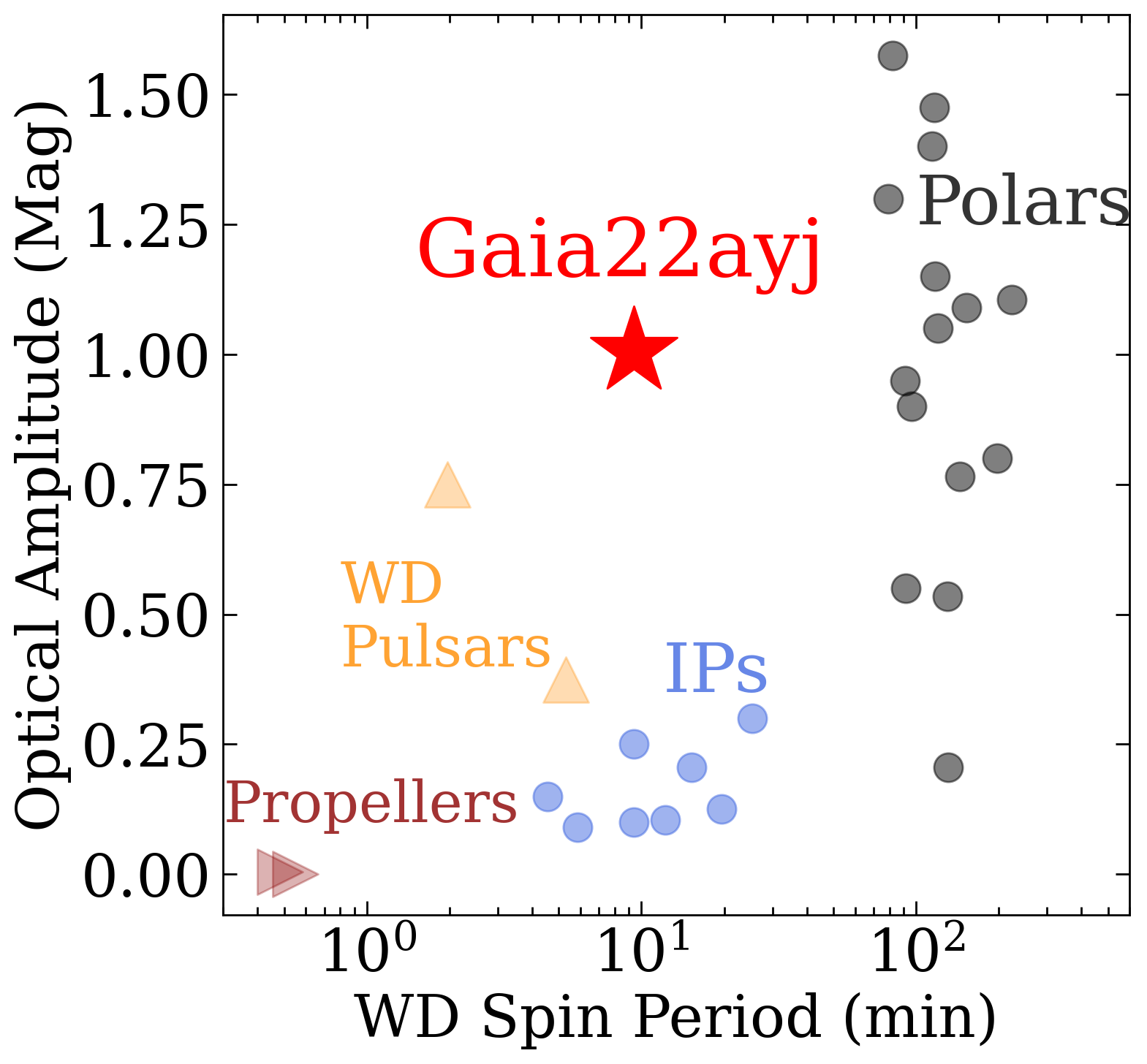}\\
    \includegraphics[width=0.45\textwidth]{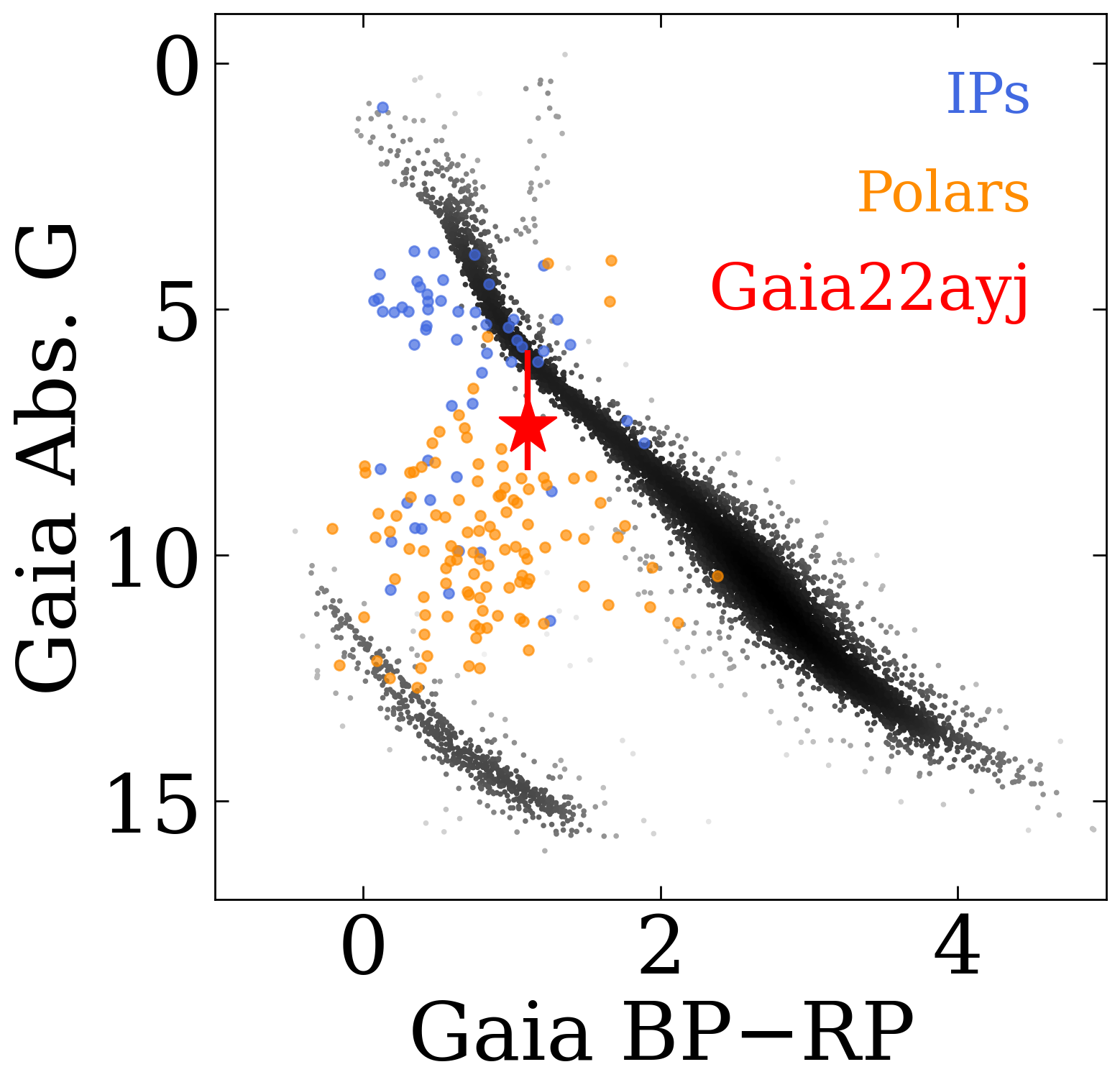}
    \caption{\textit{Top: }Gaia22ayj occupies a new region in the phase space of WD spin period vs. optical amplitude, suggesting that, at least empirically, it represents a new class of magnetic CVs. \textit{Bottom: }Gaia22ayj is roughly located between IPs and polar CVs in the \textit{Gaia} HR diagram.}
    \label{fig:phase-space}
\end{figure}

Gaia22ayj is observationally a new class of object. It occupies a new region in the phase space of photometric period (spin period) vs. optical amplitude. To empirically show that Gaia22ayj represents a new subclass of CVs, Figure \ref{fig:phase-space} situates it in the phase space of WD spin period vs. peak-to-peak optical amplitude for various magnetic WD subtypes. The vertical axis shows the amplitude in ZTF $r$ magnitudes for both known WD propellers, both known WD pulsars, a representative sample of IPs taken from the ``ironclad list" compiled by Koji Mukai\footnote{\url{https://asd.gsfc.nasa.gov/Koji.Mukai/iphome/catalog/alpha.html}}, and a representative sample of polar CVs from the Ritter and Kolb catalog \citep{2003ritterkolb}. Finally, we show Gaia22ayj on a \textit{Gaia} HR diagram (100-pc sample) compared to polars and IPs from the catalog of \citet{2020abril}, with only systems that have a three $\sigma$ measurement of parallax shown. Gaia22ayj appears to fit right between IPs and polars, owing to its likely long (but still unknown) orbital period and high accretion luminosity.

\subsection{Gaia22ayj: A Link Between White Dwarf Pulsars and Polars}
\label{sec:link}

The evidence presented thus far suggests that Gaia22ayj is an accreting analog of WD pulsars, since 1) Gaia22ayj is rapidly spinning down, being the only system with a $P/\dot{P}$ comparable to that of AR Sco, 2) has high levels of linear polarization matched only by AR Sco, and 3) is accreting, with a WD spin period slightly longer than that of the known WD pulsars, but not yet synchronized with the orbital period as would be the case in polar CVs. 

In considering the radiation mechanisms driving the optical modulations, it is notable that in AR Sco–type systems, the optical and polarimetric signals are largely powered by synchrotron processes \citep{2017buckley}, whereas in polars and intermediate polars (IPs), cyclotron emission from the white dwarf's accretion shock is dominant \citep[e.g.][]{2017mukai}. The high levels of linear polarization in Gaia22ayj appear more consistent with AR Sco–type behavior; however, one important caveat is the observed anti-correlation between the linearly polarized pulse and the photometric pulse, which contrasts with the behavior typically seen in AR Sco–type systems \citep{2017buckley}. Circular polarization and further examination of the polarized flux in Gaia22ayj would help clarify how much of its polarized variability might be driven by unpolarized variability.

Based on our measured value of $\dot{P}$, the WD in Gaia22ayj will slow down to match the 1.3--4 h orbital periods of polars in $\approx40$ Myr, assuming a constant $\dot{P}$, though it may do so more rapidly if higher order derivatives of the spin period are measured. Curiously, this timescale appears to agree with predictions made by \cite{2021schreiber}, as seen in Supplementary Material Figure 2, where accretion commences as the WD spins down.

\begin{figure*}
    \centering
    \includegraphics[width=\textwidth]{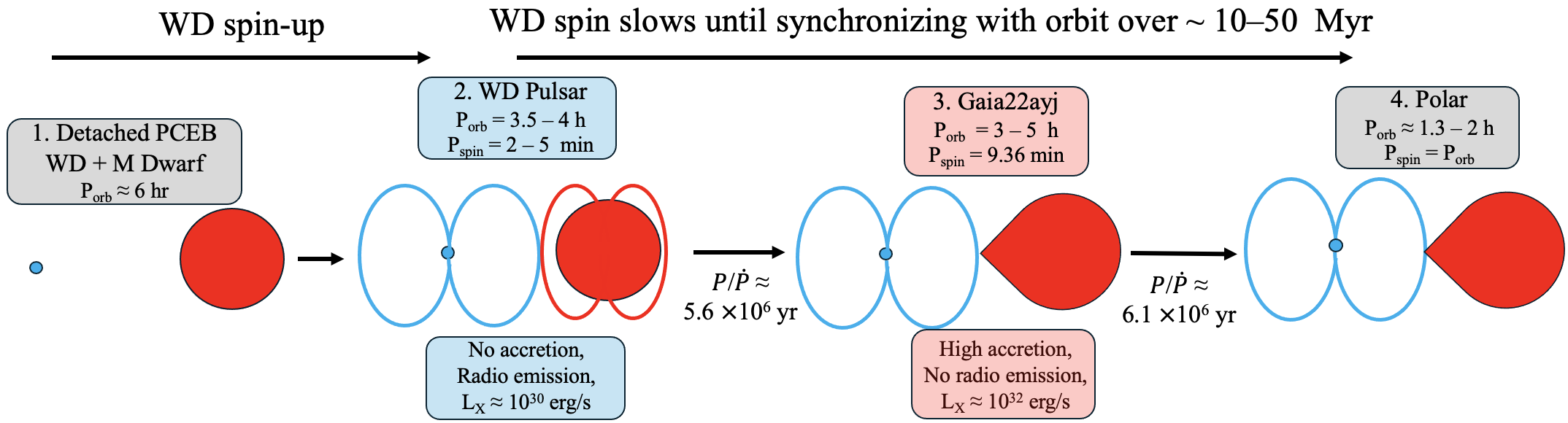}
    \caption{Cartoon of the possible evolution of WD pulsars into Gaia22ayj and then into polars. WD pulsars must be products of common envelope evolution, and WDs are likely spun up by an early accretion phase. WD pulsars are detached (non-accreting) systems, which are spinning down. Along the spin-down phase, the donor fills its Roche lobe and accretion begins, resembling Gaia22ayj. In $\approx40$ Myr, Gaia22ayj will spin down to the point where the WD spin is synchronized with the orbit, leading to the creation of a polar CV.}
    \label{fig:cartoon}
\end{figure*}

In a similar style to the evolutionary model put forth by \cite{2021schreiber}, in Figure \ref{fig:cartoon}, we present a cartoon of the evolution of Gaia22ayj. First, a WD + M dwarf system emerges out of a common envelope phase with an unknown orbital period, presumably longer than 6 h \citep{2011knigge}. Then, the WD is spun up to 2--5 min through accretion until the binary system reaches orbital periods of 3.5--4 h. At this point, the binary is detached, and a strong WD magnetic field is present, which interacts with that of the donor star, leading to pulsed radio emission. Because of the lack of accretion, X-rays are weak. As the WD spins down to $\approx$ 10 min, the donor star fills its Roche lobe and accretion commences. This leads to no radio emission (or at least much weaker emission than that of WD pulsars), and X-rays two orders of magnitude stronger than in the detached phase, over the course of $\approx40$ Myr.

\subsubsection{Is Gaia22ayj an Intermediate Polar?}
The arguments put forth above suggest that Gaia22ayj is a new class of accreting magnetic WDs, but which could also be considered a new subtype of intermediate polars (IPs). The WD spin period, X-ray luminosity, and possible orbital period are all similar to the majority of IPs \citep[e.g.][]{2019suleimanov}. However, the extreme levels of linear polarization (Figure \ref{fig:hipercam}), high amplitude optical modulation (Figure \ref{fig:phase-space}), and rapid spin-down rate (Figure \ref{fig:pdot_space}), are completely unlike any known IP. As more analogous systems are discovered, we suggest usage of the term ``Gaia22ayj-like'' systems. When an indisputable link between WD pulsars and polars is found, ``post-pulsar'' could be a suitable term since ``pre-polars'' already exist.

\subsection{The Biggest Unknowns: Orbital Period and Magnetic Field Strength of Gaia22ayj}

\begin{figure}
    \centering
    \includegraphics[width=0.5\textwidth]{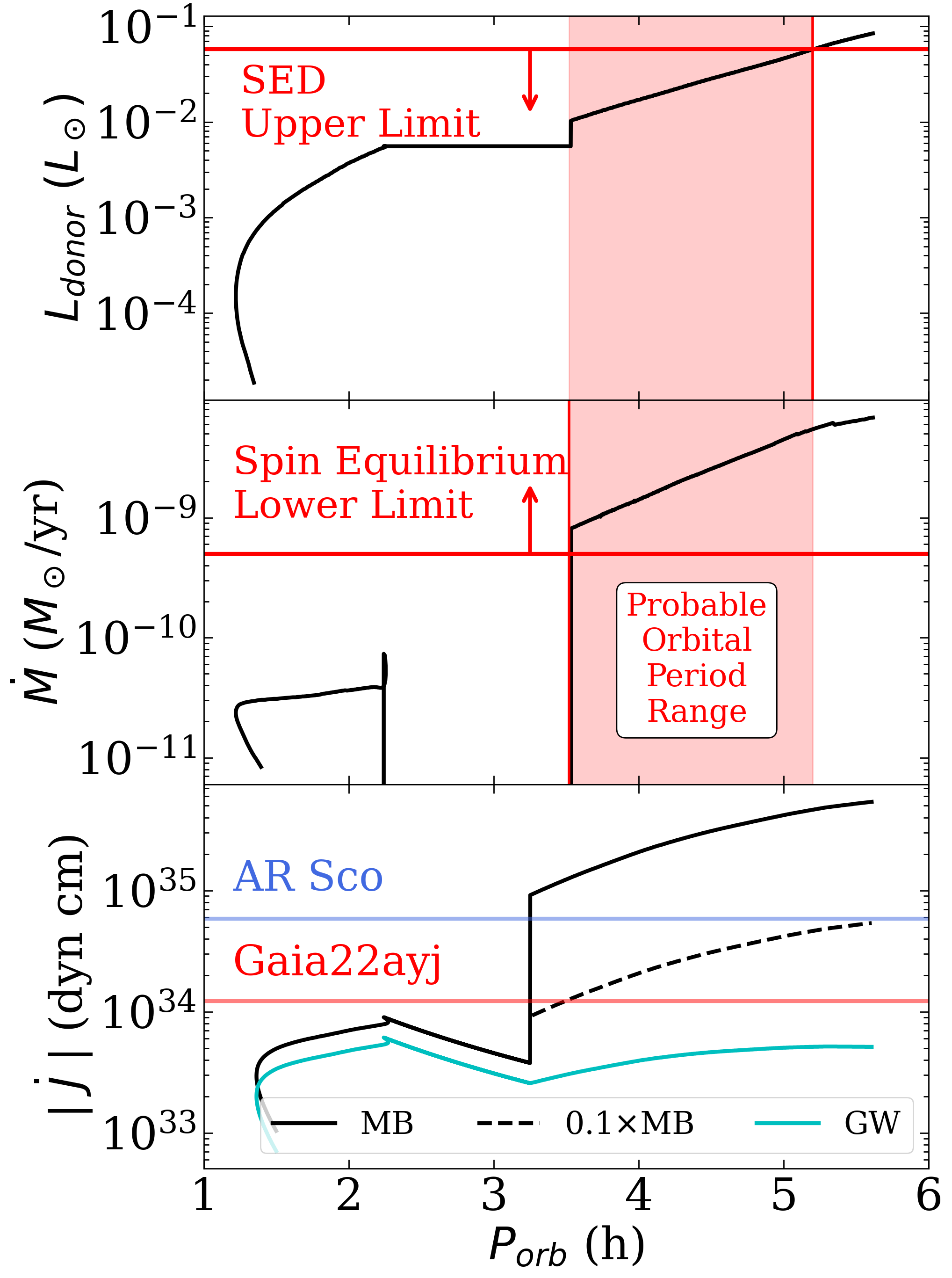}
    \caption{The evolutionary models of \cite{2011knigge} illustrate the possible orbital period of Gaia22ayj. The SED at light curve minimum (Figure \ref{fig:sed}) places upper limits on the donor luminosity (\textit{top}). The fact that Gaia22ayj is accreting and not flinging material out as a ``propeller" sets a lower limit on the accretion rate (\textit{middle}). Combined, they constrain $P_\textrm{orb} = 3.5-5.2$\,h, though future observations that detect donor lines and measure RVs are needed to test this. \textit{Bottom:} Even assuming weakened (10\%) magnetic braking compared to the \cite{2011knigge} models, the angular momentum transferred by Gaia22ayj back to the orbit (assuming constant spin-down) will not detach the binary.}
    \label{fig:evolution}
\end{figure}

At this point, the biggest question surrounding Gaia22ayj is: what is its orbital period? The known WD pulsars have orbital periods of 3.5 -- 4 h, while the majority of polars have orbital periods between 1.3 -- 4 h. Therefore, for Gaia22ayj to be a possible link between the two, it should have an orbital period in the $\approx$ 3--4 h range. In Figure \ref{fig:evolution}, we show the CV evolutionary tracks by \cite{2011knigge} as a function of orbital period. In the upper panel, we plot the upper limit of the donor luminosity derived from fitting the SED at light curve minimum (Figure \ref{fig:sed}). If a donor star exceeded this luminosity, it would definitely be seen in the optical spectra of Gaia22ayj. This allows us to place an upper limit on the orbital period of $\approx5.2$ h. In the middle panel of that figure, we place a lower limit based on the assumption that the WD in Gaia22ayj is in spin equilibrium. Because we know the WD spin period and have a sense of the magnetic field strength, we place lower limits on the accretion rate (Figure \ref{fig:equilibrium}). Plotting this value of $\dot{M} = 5\times 10^{-10} M_\odot\; \textrm{yr}^{-1}$ on the CV evolutionary tracks, we obtain an orbital period lower limit of $\approx3.5$ h. We infer that Gaia22ayj could have an orbital period in the range 3.5--5.2 h, though it could be in the CV orbital period gap with an even lower orbital period. Because there is overlap with the periods of the known WD pulsars, we conclude that given the current data, it is likely that Gaia22ayj could indeed be a link between WD pulsars and polars.

In the lower panel of Figure \ref{fig:evolution}, we show the absolute value of the derivative of angular momentum, $|\dot{J}|$, as a function of orbital period, assuming the \cite{2011knigge} CV evolutionary tracks. The black and cyan curves represent the contribution to angular momentum loss from magnetic braking and gravitational wave radiation, respectively. We show a weakened magnetic braking prescription, simply reduced by a factor of ten, since it has been suggested that magnetic braking models in the past have been too strong \citep[e.g.][]{2020belloni, 2022elbadry}. As Gaia22ayj spins down to match the orbital period of polars, the angular momentum should be transferred to the orbit, which could lead to the system becoming detached. We plot horizontal lines which represent the change in angular momentum transferred to the orbit over 40 Myr timescales, starting from the spin periods of AR Sco (1.97 min, blue) and Gaia22ayj (9.36 min, red), and ending at 2 h. Effectively for all plausible orbital periods of Gaia22ayj (3.5 -- 5.2 h), the red line is below both the black and dotted black lines, which suggests that the angular momentum transferred to the orbit during the spin-down phase will not lead to the binary detaching. However, weaker magnetic braking prescriptions could lead to detachment and should be explored in further detail.

Returning to the main question here---how can the orbital period of Gaia22ayj be detected? We have shown that $\approx2$ h on 10-m class telescope time, unfortunately, is inadequate. Near-infrared spectroscopy with similar class telescopes is another possible option, and should be undertaken in upcoming years. Clearer, low harmonic cyclotron humps should be present in that regime, and it is possible that donor lines (either emission or absorption) will stand out more clearly. Ultraviolet spectroscopy with \textit{HST} may detect the RV shift of the WD with respect to the center of mass, though since this is a long period (few hours) system, this effect is likely to be minimal. The discovery of Gaia22ayj analogs that may exhibit clearer donor lines is likely the best prospect.

X-ray timing of Gaia22ayj, if it confirms a different period than that measured in the optical, may also help solve this mystery. Assuming that the 9.36-min optical period is the WD spin-orbit beat and the 9.64-min X-ray period is the WD spin, we find that $P_\textrm{orb} = 5.37$ h, in rough agreement with the limits we outline in Figure \ref{fig:evolution}. Because of the limitations of the \textit{Swift} GTI in our analysis, however, we can at best report this as a candidate orbital period. An analogy can be drawn to the prototypical diskless IP, V2400 Oph, where an orbital period has never been directly measured, but instead inferred from the WD spin (seen in circular polarization) and spin-orbit beat (seen in optical photometry) \citep{1995buckley}. 

Finally, we emphasize that we have not provided a true measurement of the magnetic field strength of Gaia22ayj due to the lack of visible cyclotron harmonics. Future work modeling the optical light curve and/or optical polarimetry, as well as near-infrared spectroscopy to find lower order (stronger) harmonics could shed light on the true value of the magnetic field strength.

\subsection{Spin-down Rate of Gaia22ayj, Duration of Evolutionary Phase, and Rarity of Similar Systems}
Another big question remains: if Gaia22ayj represents an evolutionary phase between WD pulsars and polar CVs, why is it so much more distant (2.5 kpc) than the two known WD pulsars (116 and 237 pc)? This means that Gaia22ayj-like systems are 1) either rare outcomes or 2) represent very short-lived phases of CV evolution. To address the first point, the main evolutionary model of \cite{2021schreiber} argued that WD pulsars first spin down to synchronicity with the orbit, commence wind accretion as low accretion rate polars, and then become true, accreting polars. This would mean that Gaia22ayj systems would just represent rare cases in which Roche-lobe filling occurs before WD spin-orbit synchronicity. Nevertheless, Figure 2 in the Supplementary Material of \cite{2021schreiber} did predict a similar synchronization timescale with ongoing accretion, exactly as is seen in Gaia22ayj. To address the second point, it could be that $\dot{P}$ keeps increasing until WD-spin orbit synchronicity is nearly reached. This is supported by the fact that $\dot{P}$ is a factor of $\approx4$ higher in Gaia22ayj than in AR Sco. Ongoing high-speed photometric campaigns of the known WD pulsars and Gaia22ayj are highly encouraged to find evidence for higher order spin period derivatives which could probe this idea.

\subsection{Gaia22ayj-like Systems in the Rubin Legacy Survey of Space and Time (LSST)}
The LSST is expected to reach $\approx24$ mag, a factor of 16 deeper than ZTF in flux \citep{2019rubin}. This means that the LSST will be able to detect a system of a given luminosity four times farther away than the farthest system detectable by ZTF. Assuming that Gaia22ayj-like systems are young and concentrated in the disk, at the very minimum, the LSST will detect dozens more. Based on its extraordinary optical light curve amplitude, we expect that Gaia22ayj-like systems will stand out in LSST, even early on in the 10-year survey. Polars also have high amplitudes, but period searches of LSST data will distinguish them from Gaia22ayj as we have done here with ZTF data. 

\begin{figure}
    \centering
    \includegraphics[width=0.5\textwidth]{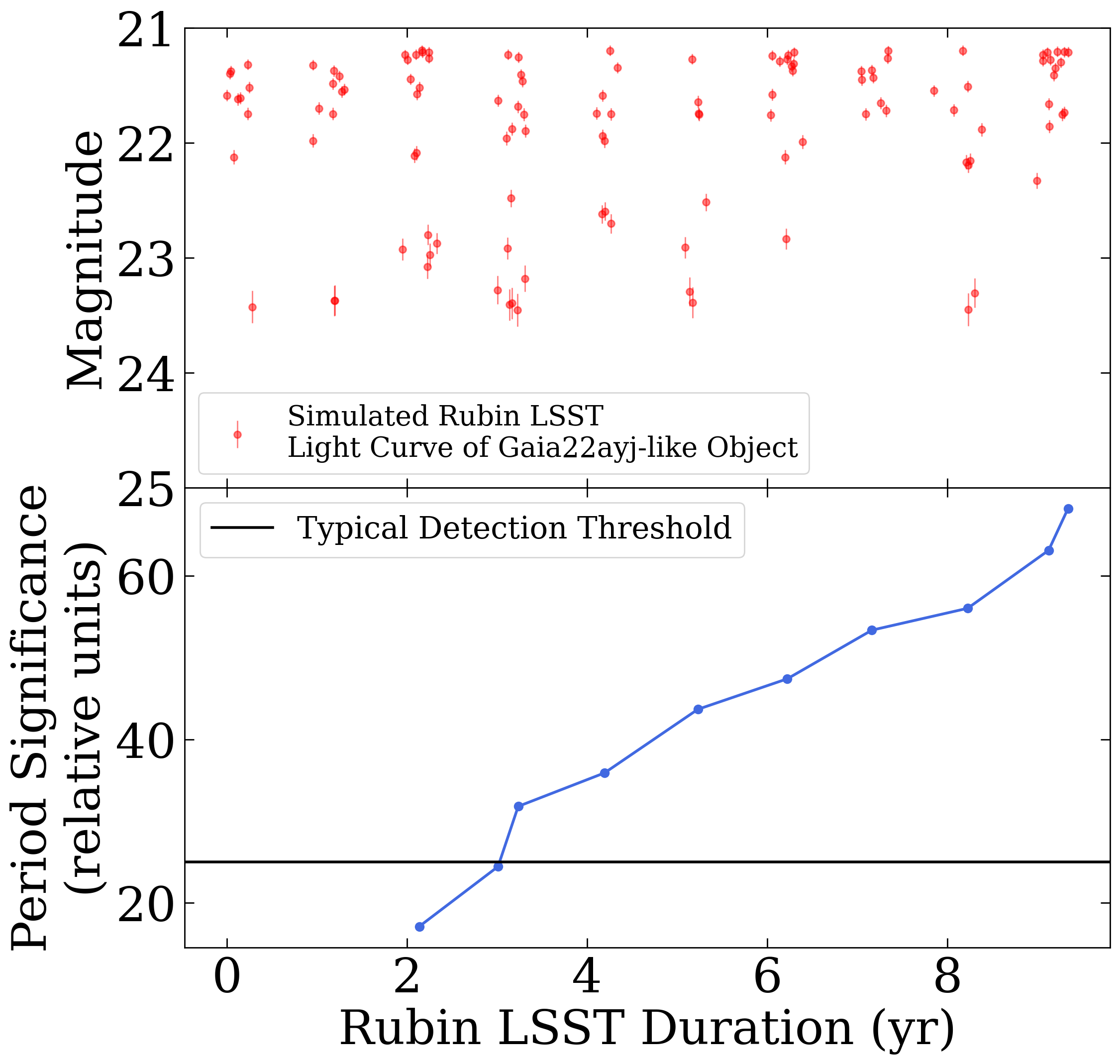}
    \caption{Simulated $r$-band light curve by injecting a Gaia22ayj-like signal into the simulated cadence of the LSST (\textit{top}). Given typical periodicity detection thresholds, tested on real data with ZTF, Gaia22ayj-like systems should be detectable $\approx3$ yr after the start of the LSST, though combining data in multiple filters, such a signal could be detected earlier.}
    \label{fig:rubin}
\end{figure}

Taking the simulated ten-year LSST baseline cadence (v3.4) described in \cite{2022lsst_opsim2}, which is based on the OpSim\footnote{\url{https://github.com/lsst/rubin_sim}} tool outlined by \cite{2016lsst_optsim} (available from the following URL: \url{https://s3df.slac.stanford.edu/data/rubin/sim-data/sims_featureScheduler_runs3.4/baseline/}), we injected a Gaia22ayj-like signal at the actual position of Gaia22ayj, but $\approx3$ mag fainter than was observed by ZTF, with error bars scaled from typical ZTF values near the detection limit. In Figure \ref{fig:rubin}, we show the simulated $r$-band LSST light curve of such a system. We also show the period significance as a function of time elapsed since the start of the LSST, calculated by taking the peak of the Lomb-Scargle periodogram divided by the median absolute deviation. We find that after $\approx3$ years into the LSST, such a signal could be detected above a significance threshold of 25. Given the significance defined above, 25 is the typical 95\% percentile for ZTF light curves (i.e. only 95 percent of ZTF light curves have such a high periodicity significance). We conclude that while Gaia22ayj-like signals could be reliably detected in the $r$ band in LSST within three years of survey start, it could be possible that through combining multiple filters, such objects could be discovered earlier. A similar claim was made and alternative observing strategies were proposed for other rapidly evolving astronomical phenomena in \cite{2022bellm}, suggesting that a higher LSST cadence could be beneficial for a multitide of science cases.

\section{Conclusion}
\label{sec:conclusion}

Gaia22ayj is a remarkable accreting magnetic WD that empirically represents a new class of magnetic CVs, and could be the missing link between WD pulsars and polar CVs. At the very minimum, Gaia22ayj:
\begin{enumerate}
    \item Pulses at optical and near-IR wavelengths on a 9.36-min period. High-speed optical photometry over six years reveals that it is slowing down, with $\dot{P} = (2.89\pm 0.12)\times 10^{-12}\; \textrm{s s}^{-1}$ and $P/\dot{P} \approx 6.1^{+0.3}_{-0.2}\times 10^6$ yr.
    \item Pulses at extreme levels, changing in brightness by a factor of 7.5--10 at optical wavelengths within the span of 2.5 minutes. Broadband spectral modulation, reminiscent of cyclotron emission in polars, is responsible for this.
    \item Likely hosts a magnetic WD accreting from a Roche-lobe-filling donor. Linear polarization levels reaching 40\% are a clear sign of magnetism. An outburst seen by ZTF, ATLAS, and \textit{Gaia} resembling outbursts from IPs, as well as broad ($v\sin i \approx 1200 \textrm{ km s}^{-1}$) double-peaked Balmer, He\,{\sc i}, and He\,{\sc ii} emission lines indicate ongoing accretion.
    \item Is a luminous X-ray source, with $L_X = 2.7_{-0.8}^{+6.2}\times10^{32} \textrm{ erg s}^{-1}$ in the 0.3--8 keV band, comparable to most IPs and the most luminous polars. It is not detected in the radio with a 3$\sigma$ upper limit of 15\,$\mu$Jy.
    \item Does not show any emission-line RV shifts over the course of $\approx2$ h of observations with Keck I/LRIS. Donor lines are only marginally detected after averaging multiple spectra and do not shift on these timescales, impeding a measurement of the orbital period. 
\end{enumerate}

Based on these properties, we argued that Gaia22ayj could be a link between WD pulsars and polars, for the following reasons:
\begin{enumerate}
    \item Gaia22ayj is spinning down at nearly the same characteristic timescale as AR Sco, $\sim 5\times 10^6\,\textrm{yr}$. 
    \item Gaia22ayj shows extreme levels of linear polarization (40\%) only seen in AR Sco, and not in either IPs or polars.
    \item The possible spin period of Gaia22ayj (9.36 min) is longer than that of AR Sco (1.97 min) and J1912 (5.3 min). Because accretion is ongoing, and the WD in Gaia22ayj is spinning down, in $\approx40$ Myr it will evolve into a polar, where the WD spin is locked with the orbit (assuming constant spin-down).
    \item The orbital period is indirectly constrained, through the nondetection of the donor star and spin equilibrium arguments, to be in the 3.5--5.2 h range, overlapping with the orbital periods of AR Sco (3.6 h) and J1912 (4.0 h).
\end{enumerate}

Other candidates (high brightness amplitude variables with $\sim$ 10-min periods) are emerging from current releases of SRG/eROSITA and ZTF, compiled using the methods outlined in \cite{2024rodriguez_diagram} and surveys presented in \cite{2024rodriguez_erosita}. Future, deeper releases from SRG/eROSITA \citep{2021predehl, 2021sunyaev} will likely reveal several more systems, though the Rubin LSST \citep{2002lsst, 2019rubin} will enable the optical discovery of at least dozens more systems within the first few years of the start of the survey. 

The detailed characterization of Gaia22ayj demonstrates that there were previously missing details regarding the true diversity of magnetic WD binaries. This shows that rare systems could represent short-lived intermediate stages of binary evolution and lead to an improved understanding of magnetic field generation in WD binaries. The discovery of exotic systems like Gaia22ayj could hint at other intermediate stages of magnetic WD systems, which have been proposed to be responsible for the emerging class of long-period radio transients (LPTs). Multiwavelength (radio, X-ray, optical) surveys and follow-up efforts show promise in revealing such systems over the next decade.

\section{Acknowledgments}

We wish to dedicate this work to the memory of our colleague and friend Tom Marsh. Tom's enthusiasm to work on this object and rapid efforts to facilitate data collection truly made this project possible. 

ACR acknowledges support from an NSF Graduate Fellowship. ACR thanks the LSST-DA Data Science Fellowship Program, which is funded by LSST-DA, the Brinson Foundation, and the Moore Foundation; his participation in the program has benefited this work. PR–G acknowledges support by the Spanish Agencia Estatal de Investigación del Ministerio de Ciencia e Innovación (MCIN/AEI) and the European Regional Development Fund (ERDF) under grant
PID2021--124879NB--I00.
MRS is supported by FONDECYT (grant numbers 1221059) and eRO-STEP (SA 2131/15-2 project number 414059771).

Based on observations made with the Gran Telescopio Canarias (GTC), installed at the Spanish Observatorio del Roque de los Muchachos of the Instituto de Astrofísica de Canarias, on the island of La Palma. Based on observations obtained with the Samuel Oschin Telescope 48-inch and the 60-inch Telescope at the Palomar Observatory as part of the Zwicky Transient Facility project. ZTF is supported by the National Science Foundation under Grants No. AST-1440341 and AST-2034437 and a collaboration including current partners Caltech, IPAC, the Weizmann Institute of Science, the Oskar Klein Center at Stockholm University, the University of Maryland, Deutsches Elektronen-Synchrotron and Humboldt University, the TANGO Consortium of Taiwan, the University of Wisconsin at Milwaukee, Trinity College Dublin, Lawrence Livermore National Laboratories, IN2P3, University of Warwick, Ruhr University Bochum, Northwestern University and former partners the University of Washington, Los Alamos National Laboratories, and Lawrence Berkeley National Laboratories. Operations are conducted by COO, IPAC, and UW. 

Some of the data presented herein were obtained at Keck Observatory, which is a private 501(c)3 non-profit organization operated as a scientific partnership among the California Institute of Technology, the University of California, and the National Aeronautics and Space Administration. The Observatory was made possible by the generous financial support of the W. M. Keck Foundation. 
The authors wish to recognize and acknowledge the very significant cultural role and reverence that the summit of Maunakea has always had within the Native Hawaiian community. We are most fortunate to have the opportunity to conduct observations from this mountain. We are also grateful to the staff of Palomar Observatory for their assistance in carrying out observations used in this work.

Partly based on observations made with the Nordic Optical Telescope, owned in collaboration by the University of Turku and Aarhus University, and operated jointly by Aarhus University, the University of Turku and the University of Oslo, representing Denmark, Finland and Norway, the University of Iceland and Stockholm University at the Observatorio del Roque de los Muchachos, La Palma, Spain, of the Instituto de Astrofisica de Canarias.
The data presented here were obtained with ALFOSC, which is provided by the Instituto de Astrofisica de Andalucia (IAA) under a joint agreement with the University of Copenhagen and NOT. The observation with the Southern African Large Telescope (SALT) was obtained under program  2021-2-LSP-001 (PI: D. Buckley). Polish participation in SALT is funded by grant No. MEiN nr 2021/WK/01. DAHB acknowledges support from the National Research Foundation.

This work presents results from the European Space Agency (ESA) space mission Gaia. Gaia data are being processed by the Gaia Data Processing and Analysis Consortium (DPAC). Funding for the DPAC is provided by national institutions, in particular the institutions participating in the Gaia MultiLateral Agreement (MLA). The Gaia mission website is \url{https://www.cosmos.esa.int/gaia}. The Gaia archive website is \url{https://archives.esac.esa.int/gaia}. This work made use of data supplied by the UK Swift Science Data Centre at the University of Leicester

ECB and JK acknowledge support from the DIRAC Institute in the Department of Astronomy at the University of Washington. The DIRAC Institute is supported through generous gifts from the Charles
and Lisa Simonyi Fund for Arts and Sciences, and the Washington Research Foundation.

\bibliography{main}{}
\bibliographystyle{aasjournal}

\appendix
\section{Full High-speed Optical Photometry}
\label{sec:appendix}

\begin{figure}
    \centering
    \includegraphics[width=0.8\textwidth]{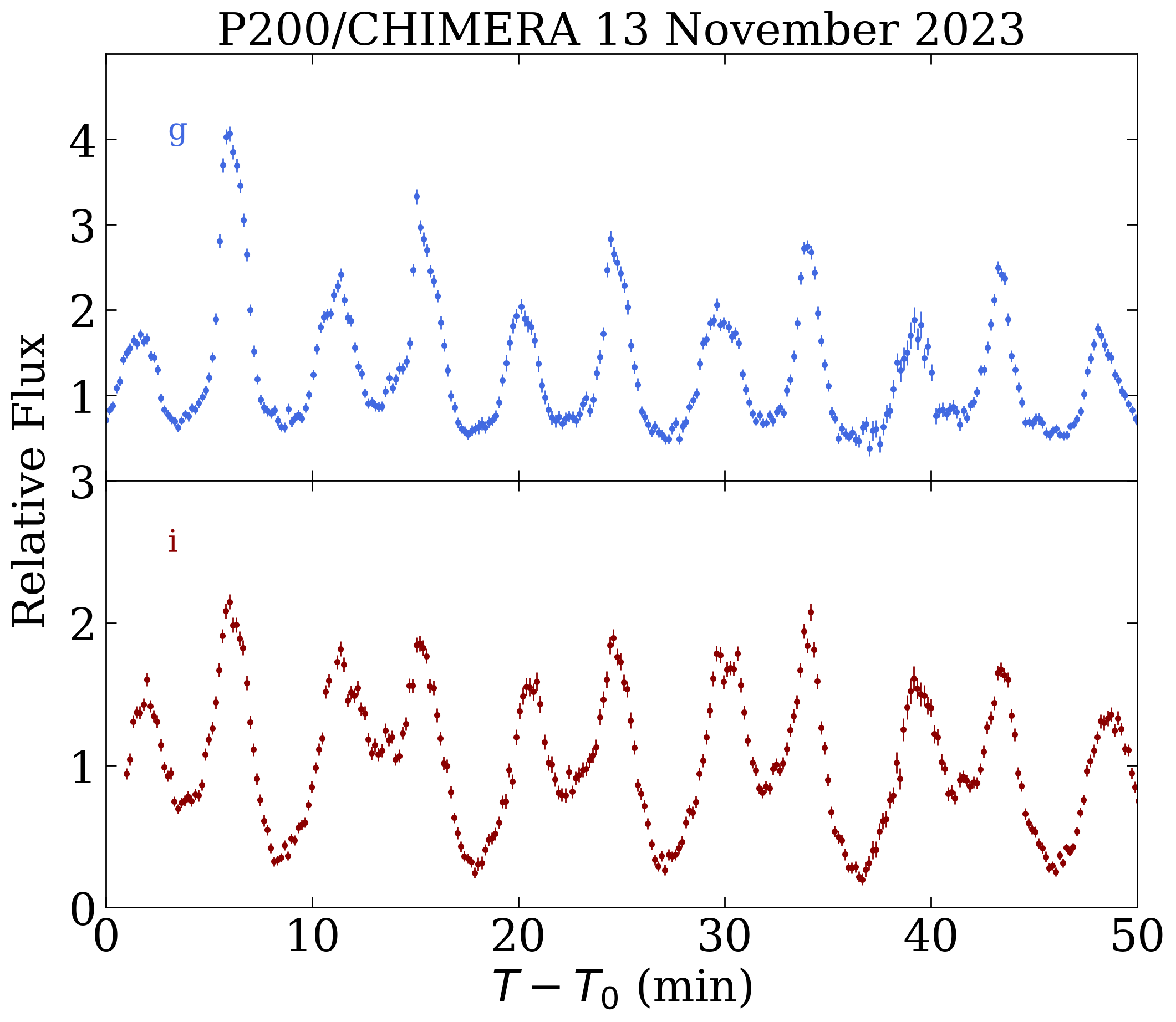}
    \caption{P200/CHIMERA light curve from 13 Nov 2023. In $g$ band, the second peak of the spin phase (higher of the two peaks) steadily decreases over the course of the observing window, from a relative flux of 4 at minute 7 to a relative flux of 2.5 at minute 42. Similar behavior is seen in the ULTRACAM light curve (Figure \ref{fig:ultracam}).}
    \label{fig:chimera}
\end{figure}

\begin{figure}
    \centering
    \includegraphics[width=0.8\textwidth]{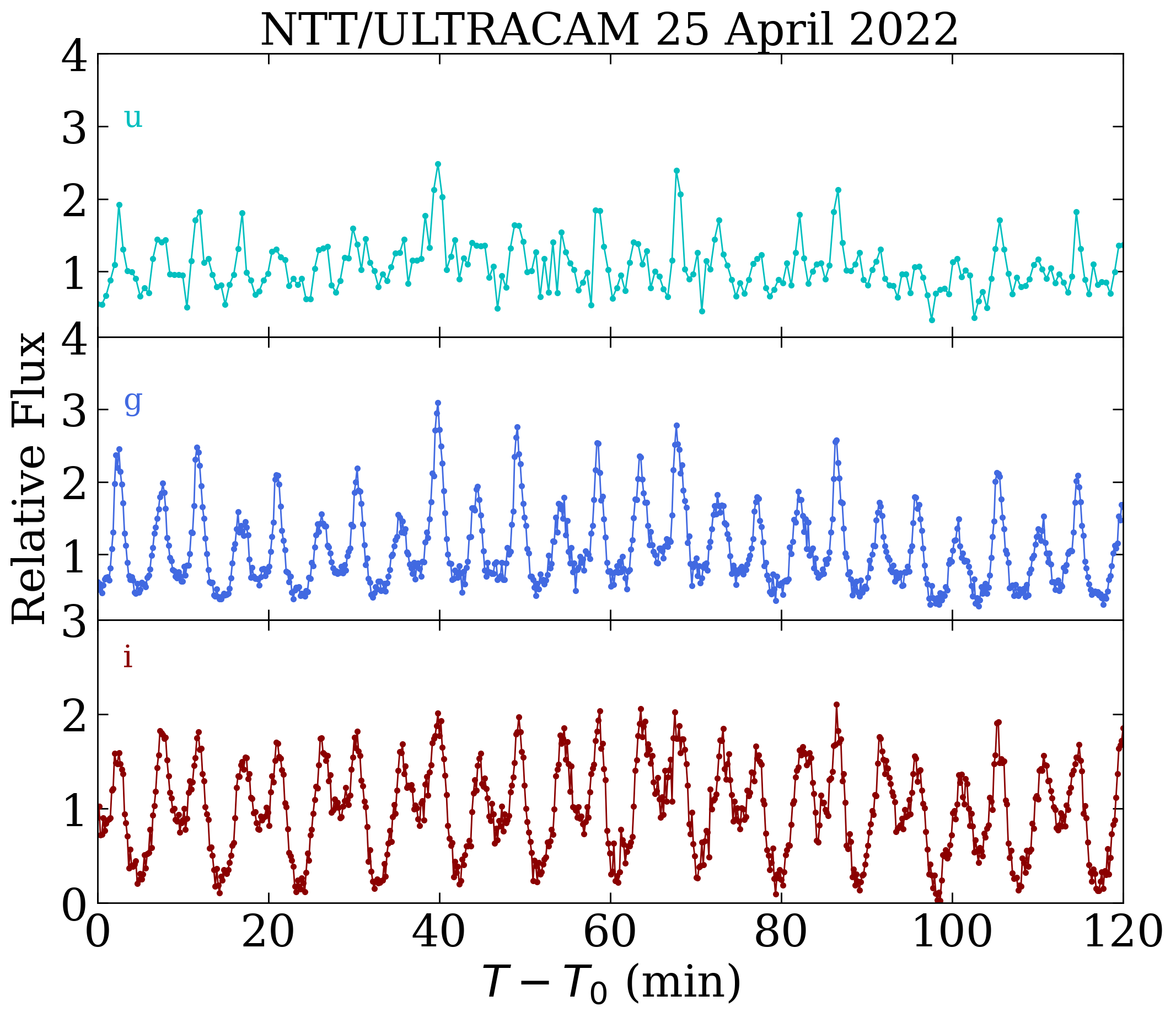}
    \caption{NTT/ULTRACAM light curve from 25 Apr 2022. In $g$ band, the second peak of the spin phase (higher of the two peaks) is variable throughout the observing window, reaching its highest value at the 40 minute mark.}
    \label{fig:ultracam}
\end{figure}

\begin{figure}
    \centering
    \includegraphics[width=0.8\textwidth]{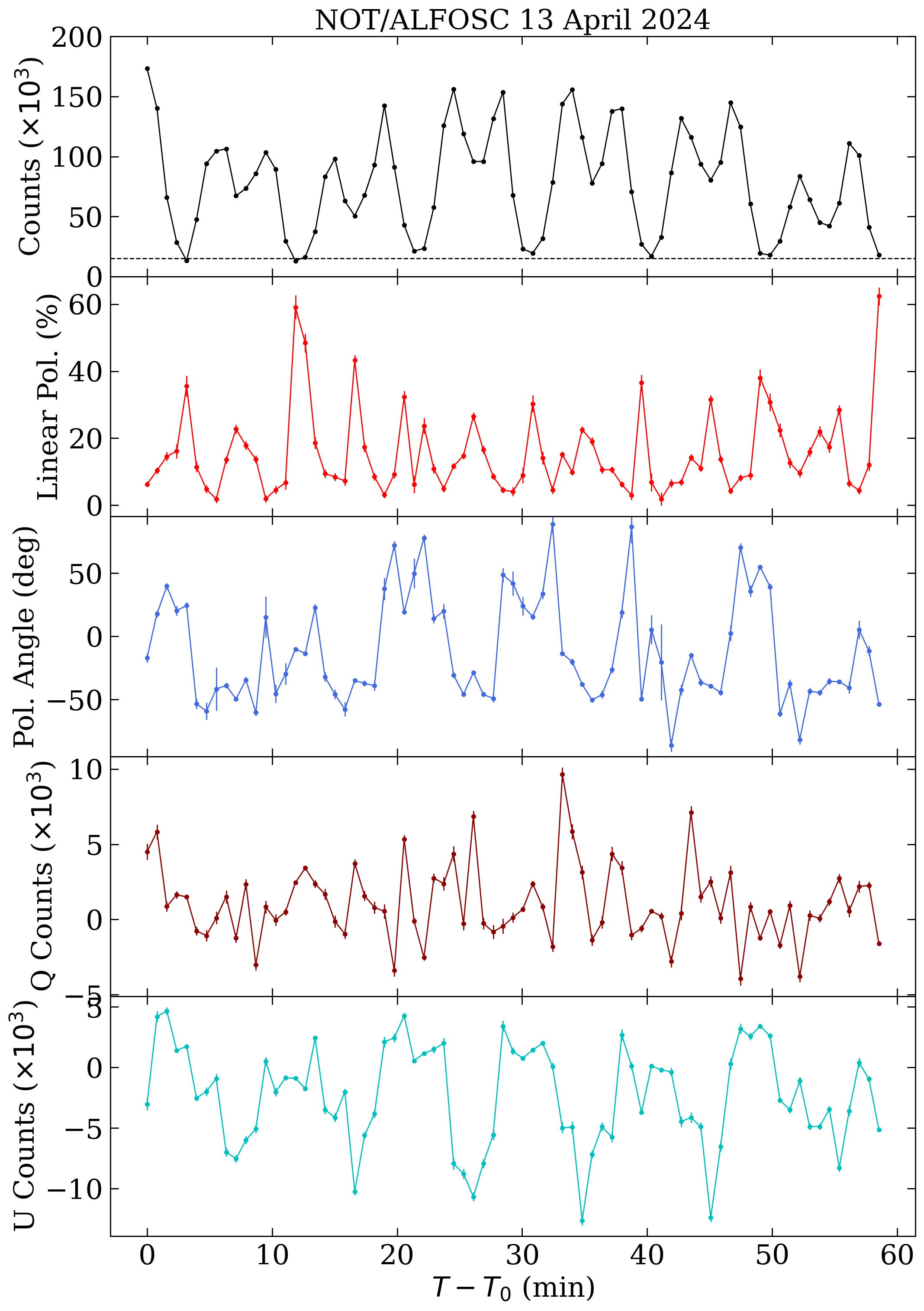}
    \caption{NOT/ALFOSC light curve from 13 Apr 2024. The dotted black line in the top panel denotes 15,000 counts, demonstrating that even at light curve minimum, a significant measurement is recorded. Linear polarization percentage regularly reaches 30\%, and exceeds 40\% in four data points. Variability in Stokes Q and U confirm that the variability in polarization percentage and angle are real and not due to noise bias. }
    \label{fig:lin_pol_tot}
\end{figure}

\begin{figure}
    \centering
    \includegraphics[width=0.8\textwidth]{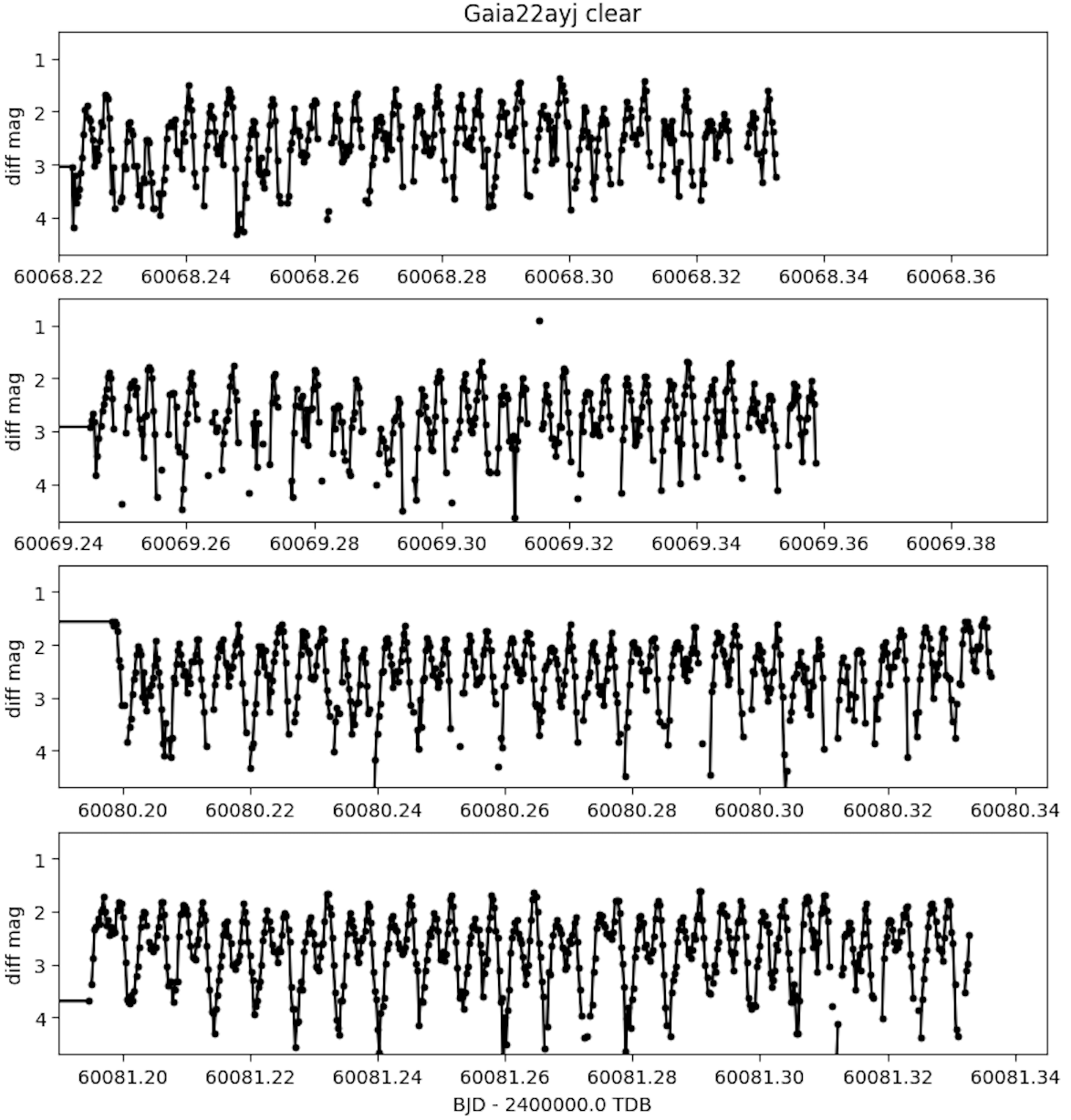}
    \caption{High speed photometry carried out with the Sutherland High Speed Optical Cameras (SHOC) on the 1m SAAO telescope reveals consistent modulation similar to that seen in other photometric runs. }
    \label{fig:saao}
\end{figure}

In Figures \ref{fig:chimera} and \ref{fig:ultracam}, we show the full high-speed light curves of Gaia22ayj from CHIMERA and ULTRACAM, respectively. Pulsations are stable in time, but not in amplitude, most noticeably in the $g$ band. The higher of the two peaks in a given spin period is particularly variable. For example, the peak near minute 7 in the CHIMERA light curve reaches a relative flux of 4, while the peak at minute 42 only reaches a relative flux of 2.5. Similar behavior is seen in the ULTRACAM light curve, where the relative flux in the higher peak reaches a relative flux of 3 at minute 40, but never does so again. In Figure \ref{fig:lin_pol_tot}, we show the full polarimetry light curve, showing that Gaia22ayj regularly reaches 30\% polarization and exceeds 40\% in four data points. In Figure \ref{fig:saao}, we show high speed photometry from a several-hour campaign carried out with the Sutherland High Speed Optical Cameras (SHOC) on the 1m SAAO telescope. 

\end{document}